\documentclass[conference]{IEEEtran}
\usepackage{url}
\usepackage{mathptmx}
\usepackage{amsmath}
\usepackage{times}
\usepackage{xcolor}
\usepackage{amsmath}
\usepackage[utf8]{inputenc}
\usepackage{epsfig,endnotes,color}
\usepackage{array}
\usepackage{multirow}
\usepackage{xspace}
\usepackage{verbatim}
\usepackage{hyperref}
\usepackage{wrapfig}
\usepackage{mathptmx}
\usepackage[english]{babel}
\usepackage{graphicx}
\usepackage{csvsimple}
\usepackage{epstopdf}
\usepackage{datetime}
\usepackage{cleveref}

\usepackage{amssymb}
\usepackage{tabularx}
\usepackage{float}

\usepackage{algorithm}
\usepackage{algorithmicx}
\usepackage{algpseudocode}

\newtheorem{definition}{Definition}

\begin{document}
\title{XYZ Privacy}

\author{
	\IEEEauthorblockN{Josh Joy, Dylan Gray, Ciaran McGoldrick, Mario Gerla}
	\IEEEauthorblockA{{\{jjoy,dylangray9,ciaran,gerla\}}@cs.ucla.edu}
}

\newcommand{\titlename}{XYZ Privacy~}

\maketitle

\begin{abstract}

Future autonomous vehicles will generate, collect, aggregate and consume significant volumes of data as key gateway devices in emerging Internet of Things scenarios.  While vehicles are widely accepted as one of the most challenging mobility contexts in which to achieve effective data communications, less attention has been paid to the privacy of data emerging from these vehicles.  The quality and usability of such privatized data will lie at the heart of future safe and efficient transportation solutions.

In this paper, we present the \titlename mechanism. \titlename is to our knowledge the first such mechanism that enables data creators to submit multiple contradictory responses to a
query, whilst preserving utility measured as the absolute error from the actual original data. The functionalities are achieved in both a scalable and secure fashion. For instance, individual location data can be
obfuscated while preserving utility, thereby enabling the scheme to transparently integrate with existing systems (e.g. Waze). A new cryptographic primitive Function Secret Sharing is used to achieve non-attributable writes and we show an order of magnitude improvement from the default implementation.

\end{abstract}

 \section{Introduction}

Researchers are becoming increasingly interested in studying smart city activities and interactions, such as pedestrians, drivers and traffic, city resources (e.g., energy) and city environment (e.g., pollution, noise). These studies are commonly based on Open Shared Data made available by several Smart City testbeds around the country. To this end, Open Data Science enables researchers to collect the data, analyze and process it with Data Mining and Machine Learning techniques and create accurate models that allow them to credibly validate smart city design methodologies.

There is a growing demand for researchers and manufacturers to deploy their technologies in real vehicles, roads and cities. Rather than requiring each stakeholder working in the area to create new solutions for securing their experimental or vehicular infrastructure, we propose a highly scalable, common ``privacy" infrastructure that enables  the non-attributable dissemination of data, whilst simultaneously conserving and preserving the critical information content properties required for value added service provision by aggregators and upstream analysts.

As smart city experiments are frequently performed on massive scale with public participants, it is prudent to surmise that some may seek to exploit the data for illicit purposes. Publicly releasing data with exact answers to queries (without sanitization) has resulted in numerous privacy violations and attacks, e.g., relating to unintentional medical data disclosure of high profile governors~\cite{Sweene02}, shutdowns of seemingly innocuous open data machine learning competitions~\cite{netflix-privacy-lawsuit}, location tracking attacks and DoS attacks~\cite{DBLP:conf/mobisys/WangWWNZZ16}, and unintentional sharing of mobility patterns of high profile US citizens with foreign governments~\cite{uber-privacy-china}. 

k-anonymity introduced by Sweeney in 1998~\cite{Sweene02} was among the first privacy techniques to address publicly releasing data in a privacy-preserving manner. Roughly speaking, k-anonymity seeks to blend a single data owner's personal attribute with at least $k$ other data owners such that the single data owner is indistinguishable from $k-1$ other data owners. For example, if a particular data owner's record reporting a particular disease is publicly released with $1000$ other data owners records with the same disease, the data owner is indistinguishable from $999$ other data owners.

However, there are known impossibility results for attempts to preserve privacy while releasing exact answers. Dinur and Nissim showed in 2003 that it is impossible to reveal exact aggregation information while simultaneously preserving privacy (against a polynomial adversary)~\cite{DBLP:conf/pods/DinurN03}. Thus, perturbation \textit{must} be injected in order to guarantee privacy (privacy defined as an adversary is unable to determine the value of a targeted individual with a probability greater than 50\%)~\cite{DBLP:conf/pods/DinurN03}. Alternatively, noiseless privacy has been proposed which does not add additional privacy noise. However, noiseless privacy requires strong adversary assumptions such as the adversary has limited or no auxiliary information and is restrictive regarding multiple queries and composability~\cite{DBLP:conf/asiacrypt/BhaskarBGLT11}. Thus, in order to have Open Shared Data, on Smart City or larger scale, there will be some notion of absolute error or distance from the ground truth due to the required perturbation. 

Differential privacy is one such privacy definition which enables realisation of this concept, and it is currently viewed as the gold standard. Roughly speaking, differential privacy says that the ability of an adversary to inflict harm should be essentially independent of whether any individual opts in to or out of, the dataset~\cite{dwork:acmarticle}. Thus, a data owner may safely utilise differential privacy techniques when sharing their personal data, as it enables them control insight into their personal information.

The Laplace mechanism satisfies differential privacy by adding privacy noise independent of the database size~\cite{DBLP:conf/tcc/DworkMNS06} by drawing privacy noise from the Laplace distribution. The Laplace mechanism is calibrated to the max difference between any two rows in the database. That is, the noise is sufficient to protect the max leakage that any particular data owner induces. For example, first a service aggregates all the data owners truthful responses. Then, the aggregation service draws from the Laplace distribution by calibrating the variance according to the desired privacy strength. Drawing from other distributions such as Gaussian also satisfies differential privacy, though the Laplace mechanism provides differential privacy and Gaussian provides ($\epsilon,\delta$)-differential privacy~\cite{DBLP:journals/fttcs/DworkR14}.

Now, consider if the Laplace mechanism is used and we desire to improve the privacy strength. The privacy strength of a given mechanism is determined by the epsilon $\epsilon$ value, which corresponds to the privacy loss measured as the ratio of the max difference between any two differing outputs. Naturally, it follows that increasing the value of $\epsilon$ adds privacy noise mitigating any utility benefits as the privacy noise increases. 

Another observation is that the use of the Laplace mechanism requires each individual to truthfully respond, relying on the output perturbation to provide privacy. This requires extra caution in the sensitive queries posed. For example, suppose we query everyone at the Brooklyn Bridge to understand how many people are currently at the Brooklyn Bridge. Regardless of the cryptographic technique or privacy mechanism used, the act of responding signals to an adversary that the data owner was indeed at the Brooklyn Bridge.

In this paper, we present the \titlename mechanism. \titlename achieves \textit{scalable} privacy. The queried population is increased beyond those merely at the Brooklyn Bridge, say to the entire New York City metropolitan area. At the same time, the absolute error is \textit{preserved} and does not increase or expand due to the  distortion of the underlying truthful population distribution (e.g., the query distorts and decreases the truthful percentage from 100\% of data owners at the Brooklyn Bridge to less than 1\% of the New York City population at the Brooklyn Bridge). Additionally, in our \titlename mechanism, data owners perform cryptographic private writes which dissociate the identifier from the data value. The aggregation operators also perform MPC to protect against malicious data owners that try to corrupt the aggregate answer.

We evaluate the scalability and accuracy of our privacy-preserving approach utilizing a vehicular crowdsourcing scenario comprising of approximately one million records. In this dataset, each vehicle reports its location utilizing the California Transportation Dataset from magnetic pavement sensors (see Section~\S\ref{sec:evaluation:accuracy}). To demonstrate the efficiency and scalability of our approach, we crowdsource and privately write 128,000 vehicle (data owner) locations in under a minute with a key size of less than 15KB, a square root reduction compared to the trivial solution whose key size would be linear in the number of data owners (i.e., the key size is the number of data owners times the message bit size). This key size reduction allows us to increase the database size to simultaneously accommodate hundreds of thousands of data owners. 

This work demonstrates, for the first time, that personal data can be crowdsourced at scale with constant error(preserving the information content of interest to upstream aggregators and analysts), strong privacy guarantees, protected with scalable cryptographic private writes, and accurately disclosed. We believe this represents an exciting new contribution to open data science, driven by the need for privacy-preserving crowdsourcing in mobile cloud contexts (e.g., traffic management).

 \section{Related Work}

~\\ \noindent \textbf{Privacy Definitions.} Differential privacy~\cite{DBLP:conf/icalp/Dwork06,DBLP:conf/tcc/DworkMNS06,DBLP:conf/eurocrypt/DworkKMMN06,DBLP:journals/fttcs/DworkR14} has been proposed as a privacy definition such that anything that can be learned if a particular data owner is added to the database could have also been learned before the data owner was added. A data owner is thus ``safe" to participate as statistical inferences amongst the aggregate are learned yet specific information regarding the individual is not learned.

However, careful consideration needs to be done when applying differentially private mechanism in practice. There is a drawback to the Laplace mechanism in graph datasets such as social networks~\cite{DBLP:conf/tcc/GehrkeLP11,DBLP:conf/crypto/GehrkeHLP12} or vehicle commuting patterns. Even if a particular data owner does \textit{not} participate, their friends that do participate leak information that can be used to deprivatize the targeted data owner (e.g., shadow profiles). For example, it is possible to learn political beliefs or sexual orientation even if a particular individual does not participate and maintain an active profile in an online social network. An adversary simply needs to analyze the similarity metrics amongst the social circles that a data owner participates in to understand politics beliefs or sexual orientation~\cite{DBLP:conf/cosn/SarigolGS14,10.1371/journal.pone.0034740,facebook-friends,Kosinski_Stillwell_Graepel_2013,FM2611}.

Furthermore, if the graph structures of the social network are eventually anonymized and released, an adversary simply needs to participate and influence the graph structure (e.g., joining a social network) to learn and influence the actual social graph before it's privatized and released. Thus, there needs to be a mechanism which also perturbs the underlying \textit{structure} of the data itself and preserves accuracy as the underlying distribution structure becomes distorted.

Sampling can be applied to weaken the associations within a graph structure. This is achieved whereby responses are randomly discarded in order to reduce the the graph dependencies leaked by a targeted individuals connections. The severed connections reduces the social circle size and makes it challenging for the adversary to make similarity inferences from reduced social circles alone. Thus, it has been shown that the strength of privacy mechanisms are increased by applying sampling and reducing the privacy leakage~\cite{DBLP:conf/stoc/NissimRS07,DBLP:conf/focs/KasiviswanathanLNRS08,DBLP:conf/ccs/LiQS12,DBLP:conf/crypto/GehrkeHLP12}.

Another popular technique which satisfies differential privacy is the randomized response mechanism, originally proposed in the 1960s~\cite{warner1965randomized,fox1986randomized}. Randomized response has been shown to be optimal in the local differentially private model~\cite{DBLP:conf/nips/DuchiWJ13} and is used by many companies today (e.g., Apple, Google~\cite{DBLP:conf/ccs/ErlingssonPK14}) due to its simplicity while satisfying the differential privacy guarantee. 

However, these protocols, such as Rappor~\cite{DBLP:conf/ccs/ErlingssonPK14}, require an inordinate amount of samples, yet still do not preserve utility. For example, even if 1 billion reports are collected, statistics from close to 1 million reports may not show up in the analysis. Thus, these type of local differentially private protocols are best suited for tracking heavy-hitters (e.g., counting the most commonly occurring elements in peaky power-law distributions)~\cite{DBLP:conf/sosp/BittauEMMRLRKTS17}. We elaborate further in Section~\ref{sec:samplingerror}.

Some protocols which leverage randomized response style mechanisms have made assumptions that the majority of the underlying truthful population truthfully responds ``Yes" (e.g., the percentage is greater than 2/3 or 3/4) in order to preserve accuracy. However, it's not clear what privacy guarantees will be provided as any adversary is able to successfully guess with greater than 50\% probability the value of any data owner in such a population. For example, suppose our query is how many home owners reside within 15 blocks from the beach, yet we ask only those home owners within 20 blocks from the beach.

Zero-knowledge privacy~\cite{DBLP:conf/tcc/GehrkeLP11} is a cryptographically influenced privacy definition that is strictly stronger than differential privacy. Crowd-blending privacy~\cite{DBLP:conf/crypto/GehrkeHLP12} is weaker than differential privacy; however, with a pre-sampling step, satisfies both differential privacy and zero-knowledge privacy. However, these mechanisms are suited for the centralized system model and rely on aggressive sampling, which significantly degrades the accuracy estimations.

Distributional privacy~\cite{DBLP:journals/jacm/BlumLR13} is a privacy mechanism which says that the released aggregate information only reveals the underlying ground truth distribution and nothing more. Each data owner is protected by the randomness of the other randomly selected data owners rather than by adding explicit privacy noise to the output. The indistinguishability from the underlying distribution protects individual data owners and is strictly stronger than differential privacy. However, it is computationally inefficient though can work over a large class of queries known as Vapnik-Chervonenkis (VC) dimension.

Other location based privacy approaches perturb a data owner's true location by randomly selecting a coordinate within some allowable range~\cite{DBLP:conf/ccs/AndresBCP13}. However, these radius perturbation based approaches face the same underlying limits as the Laplace mechanism. Increasing the allowable radius range clearly improves privacy strength, yet fundamentally the there is no underlying utility (the absolute error as measured from the ground truth).

~\\ \noindent \textbf{Sampling.} Sampling whereby a centralized aggregator randomly discards responses has been previously formulated as a mechanism to amplify privacy~\cite{DBLP:conf/crypto/ChaudhuriM06,DBLP:conf/stoc/NissimRS07,DBLP:conf/focs/KasiviswanathanLNRS08,DBLP:conf/ccs/LiQS12,DBLP:conf/crypto/GehrkeHLP12}. The intuition is that when sampling approximates the original aggregate information, an attacker is unable to distinguish when sampling is performed and which data owners are sampled. These privacy mechanisms range from sampling without a sanitization mechanism, sampling to amplify a differentially private mechanism, sampling that tolerates a bias, and even sampling a weaker privacy notion such as k-anonymity to amplify the privacy guarantees. 

However, sampling alone has several issues. First, data owners are not protected by plausible deniability as data owners do not respond ``No". Second, the estimation of the underlying truthful ``Yes" responses quickly degrades as we increase the population that truthfully responds ``No".

~\\ \noindent \textbf{Multi-party Computation.} Multi-party computation (MPC) is a secure computation model whereby parties jointly compute a function such that each party only learns the aggregate output and nothing more. However, MPC mechanisms that release the \textit{exact} answer has no strong privacy guarantees against active privacy attacks, particularly when the data is publicly published. 

A participant that does not perturb their responses and provides their \textit{exact} answer is easily attacked by an adversary that knows the values of $n-1$ participants. For example, an adversary first runs a counting query that includes all $n$ data owners and then runs a second counting query over $n-1$ data owners (the targeted data owner is the excluded row). Subtracting the two results \textit{reveals} the value of the targeted data owner. 

In contrast, the differential privacy model assumes a strong adversary that knows the $n-1$ data owner values. In this paper we combine MPC and differential privacy by introducing a sampling-based privacy mechanism that maintains constant error and show a performance optimization for a new cryptographic primitive named Function Secret Sharing~\cite{DBLP:conf/eurocrypt/BoyleGI15}.

~\\ \noindent \textbf{Private Data Upload.} Wang et al. ~\cite{DBLP:conf/nsdi/WangYGVZ17} employed and extended the function secret sharing primitive to enable efficient  private information retrieval operations that protect the data owner's queries from being learned by the database servers. They proposed an optimization by using the Matyas-Meyer-Oseas one-way compression function as an alternative to the heavy AES operations for the \textit{two party} case.  Wang et al. achieves a 2.5x speedup by utilizing one-way compression functions. However, our \titlename \textit{also} demonstrates a \textit{one order of magnitude} improvement over the default implementation of the function secret sharing protocol for the \textit{multi-party} database aggregator scenario.

~\\ \noindent \textbf{Private Stream Aggregation.} Private stream aggregation has been proposed whereby each data owner locally encrypts with their own individual key and adds differentially private noise in a distributed manner~\cite{DBLP:conf/ndss/ShiCRCS11}. However, these protocols have two fundamental concerns. First, they are not capable of operating in the mobile (e.g., vehicular) environment as the aggregator \textit{must} collect the encrypted values from each data owner. A single missing encrypted value will completely halt the protocol as the aggregator relies on the sum of the encrypted values in order to decrypt the aggregated sum. Our protocol is able to operate in the mobile environment and can safely disregard incomplete data owner uploads. Second, these protocols rely on a distributed form of Laplace noise. This means they have the same limitations as described earlier where increasing the privacy noise mitigates any potential utility. Our \titlename mechanism is able to \textit{improve} privacy while preserving utility.

 \section{Threat Model}
\label{sec:threatmodel}

The attack:  an adversary can utilize the database size (number of participants) to deduce if a particular individual is included. However, the \textit{exact} population (database) size or \textit{exact} number of participating data owners is not published or released. This mitigates auxiliary attacks whereby the adversary uses the exact counts to reconstruct the database.

The attack: an adversary can individually inspect the responses of each data owner to ascertain their truthful response. However, we select sampling probabilities less than 50\% so that an adversary does not gain an inference advantage of greater than 50\%. We also require a distributed set of aggregators whereby at least one honest aggregator does not collude with the others (e.g., a privacy watchdog like the EFF).

The attack: only a single honest data owner (or very few data owners less than $k$) participate allowing an adversary to easily deprivatize the honest data owner. However, the aggregation parties proceed in epochs where they only combine their results if at least $k$ data owners (a threshold that can be configured) participate. The honest aggregation party may refuse to share their results with the other aggregation parties thereby halting the protocol, if the number of participating data owners is below the desired threshold.

~\\
\noindent \textbf{Differential Privacy Guarantee.} The protocol we propose is a 2 round protocol. However, it is 2 rounds in the sense that 2 different values are uploaded by a single data owner. However, these values are not able to be linked to each other due to the cryptographic private write in Section~\ref{sec:privatedataupload}.

The attack: a network adversary attempts to drop the upload of the 2nd round, in order to isolate the value of a single data owner in round 1 to deprivatize the data owner and learn their truthful value. However, round 1 and round 2 are sent as a tuple $\langle round_1, round_2 \rangle$, so if any round is dropped by the network adversary both rounds are dropped.

~\\
\noindent \textbf{Pollution Attacks.} In this work we consider three different pollution attacks : 1) a malicious data owner who attempts to corrupt the private write by attempting to write to multiple rows of the database instead of a single row 2) a malicious data owner who answers a query with a single, large value in order to inflate the aggregate sum 3) a malicious data owner who repeatedly answers a query within a single epoch. More details can be found in~\ref{pollutionprotection}.

In order for the aggregators to safely accept a particular data owner's contribution, each share must be verified to be only write to a single row, i.e. the shares should resolve to unit vectors. The aggregators perform multi-party computation for each data owner in order to verify the FSS shares. We utilize recent constructions in FSS verification that do not rely on any public-key primitives~\cite{DBLP:conf/ccs/BoyleGI16}. 

The observation is the following: the dot product of any unit vector with itself is one, while the dot product of a non-unit vector is the square of the magnitude. The data owners submit blinded shares to each aggregator and then compute over the blinded shares so that the actual unit vector (i.e. data owner actual response) is not revealed. The aggregators perform an MPC to verify that the dot product of the blinded shares with itself evaluates to one, ensuring that the shares are properly formed. As long as there is at least one honest aggregation party, no aggregator learns which database row is written to. Invalid FSS shares can be quickly XORed out of the intermediate results once they are detected. Further details of the scheme can be found in the Section~\ref{sec:pollutionprotection}. \\ \section{Warm Up Construction}

Let us consider first a warm up scenario whereby each data owner privatized their truthful response utilizing the randomized response privacy mechanism and then privately uploads with an information theoretic guarantee.

First, we explain how we discretize real-numbered values to integers. We then formally describe the randomized response mechanism. Then, we describe the information-theoretic private upload mechanism.  Finally, we integrate both techniques and show the limitations, in particular, constant error is not maintained as the non-truthful population increases.  In addition, the information-theoretic private upload requires a keysize on the order of the database size making it prohibitively expensive for hundreds of thousands of data owners. We motivate the need for a more sophisticated sampling based privacy mechanism (the \titlename mechanism) and more sophisticated techniques for cryptographic compression.

\subsection{Goal}

Our primary goal is to enable large scale private querying of the population utilizing sampling mechanisms while maintaining constant error. For example, suppose we query the number of vehicles on a busy stretch of the highway. We could query only those at a particular stretch of the highway. However, we would know the stretch of the highway location of any participating data owner. Thus, the privacy protection is quite limited regardless of any perturbation performed for this particular query.

The privacy protection would be improved if we query everyone in the city. The additional data owners provide plausible deniability by increasing the potential pool of candidates that sometimes respond ``Yes" indicating they are at the particular stretch of highway. Now, if we know that someone participated in the query all we can immediately deduce is they are ``somewhere" in the city.

More generally, let $Yes_{pop}$ refer to the truthful ``Yes" fraction of the population and $No_{pop}$ refer to the truthful ``No" fraction of the population. We seek to increase the $No_{pop}$ by expanding the query to include more participating data owners. This results in lowering the percentage of the $Yes_{pop}$. Using the previous example, querying only those at the particular stretch of the highway would result in the $Yes_{pop}$ being 100\%. Expanding the query to the city would reduce this percentage to say 5\% or even 1\% or lower.

~\\
\noindent \textbf{Query.} The query is posted on the web. Data owners periodically check the web and download the query. The query is persistent and is set to expire days or weeks in the future.

\subsection{Discretization}
\label{sec:discretization}

\begin{figure}[t!]
    \includegraphics[width=1\columnwidth]{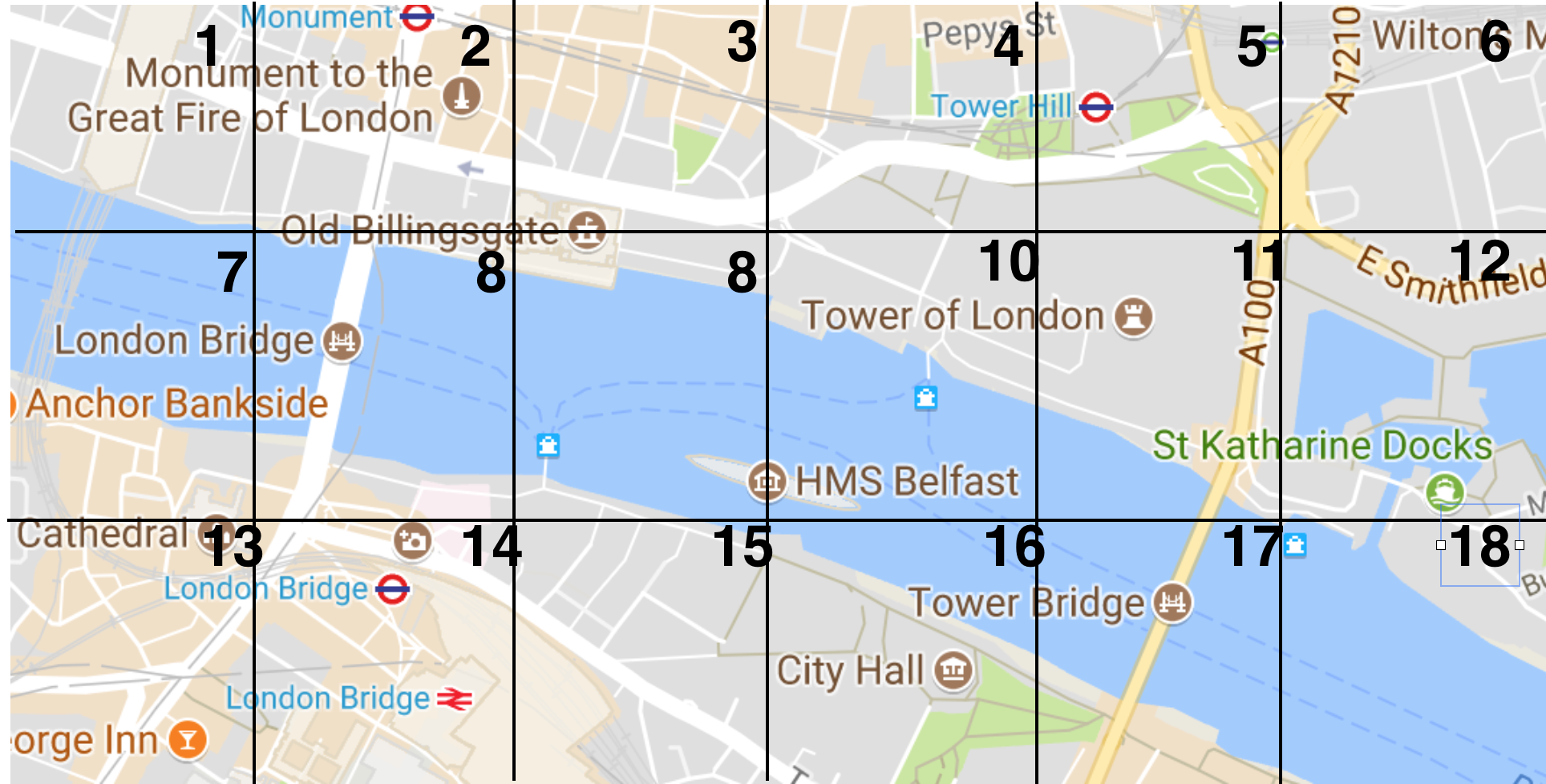}
    \caption{\textbf{Location Discretization.} Each location (latitude,longitude) is discretized to a location identifier which corresponds to a 0.25 square mile block. London Bridge corresponds to location ID 8.}
        \label{fig:discretization}
\end{figure}

We illustrate the scheme using location coordinate data, although the discretization scheme can be employed for all real valued data. Suppose a data owner currently on London Bridge participates in the protocol. First, the location is discretized to a location identifier (ID) as seen in Figure~\ref{fig:discretization}. For example, using a 16 bit identifier provides 65,536 possible locations, which covers a 64 x 64 mile square with 0.25 mile sections for a total of 4,096 square miles. For comparison Paris is 41 square miles, London is 607 square miles, New York City is 305 square miles~\cite{us-cities-area}. In Figure~\ref{fig:discretization}, London Bridge corresponds to location ID 8.

\subsection{Sampling Error}
\label{sec:samplingerror}

We now examine how the Randomized Response~\cite{warner1965randomized,fox1986randomized} mechanism grows in error as the $No_{pop}$ increases. We first formally describe the Randomized Response mechanism and then describe how the sampling error increases with the population. We will later show in Section~\ref{sec:privacymechanism} how to preserve the utility.

~ \\
\noindent \textbf{(Randomized Response)} We use two independent and biased coins. Let $\pi_1$ and $\pi_2$ refer to the heads probabilities of the first and second biased coin toss respectively. The coin toss parameters are published publicly while the number of data owners is private and needs to be estimated.

\begin{equation}
  Privatized~Value_{Yes} =
  \begin{cases}
    1 & \text{with probability } \\
    ~ & \text{$\pi_1 + (1-\pi_1) \times \pi_2$} \\
    0 & \text{otherwise}
  \end{cases}
\end{equation}

That is,  the $Yes_{pop}$ subpopulation responds ``Yes" with probability $\pi_1 + (1-\pi_1) \times \pi_2$. Otherwise they respond ``No".

\begin{equation}
  Privatized~Value_{No} =
  \begin{cases}
    1 & \text{with probability } \\
    ~ & \text{$(1-\pi_1) \times \pi_2$} \\
    0 & \text{otherwise}
  \end{cases}
\end{equation}

That is,  the $No_{pop}$ subpopulation responds ``Yes" with probability $(1-\pi_1) \times \pi_2$. Otherwise they respond ``No".

~ \\
\noindent \textbf{(Expected Value)} We now formulate the expected value in order to carry out the estimation of the underlying population. The expected value of those that respond `1' (i.e., privatized ``Yes") is the sum of the binomial distribution of each subpopulation.

\begin{multline}
\label{eq:rrexpectedvalue}
E[1] = \pi_1 \times Yes_{pop} + (1-\pi_1) \times \pi_2 \times (Yes_{pop} + No_{pop}) 
\end{multline}

~ \\
\noindent \textbf{(Estimator)} We solve for $Yes_{pop}$ by the following. Let the aggregated privatized count E[1] defined in Equation~\ref{eq:rrexpectedvalue} be denoted as $Private~Sum$. 
\begin{multline}
\label{eq:rr}
Yes_{pop} = \frac{Private~Sum - (1-\pi_1) \times \pi_2 \times (Yes_{pop} + No_{pop})}{\pi_1}
\end{multline}

That is, we first subtract from Private Sum of the ``privacy noise". We then divide by the first flip $\pi_1$ which is the sampling parameter which determines how frequently a data owner truthfully responds ``Yes" from the  $Yes_{pop}$ subpopulation.

~ \\
\noindent \textbf{(Sampling Error)} Suppose published parameters of the coin tosses are configured independently with $\pi_1=0.85$, $\pi_2=0.3$ and $100$ data owners. We estimate the underlying ``Yes" truthful population using Equation~\ref{eq:rr} by aggregating the privatized responses from all data owners, subtracting the expected value of $(1-0.85) \times 0.3 \times 100$ and dividing by $0.8$ \footnote{For instance with a 60\% truthful population, the answer to the first toss is $0.6 \times 0.85=0.51$ and the answer to the second toss is $(1-0.85) \times 0.3=0.045$}.

However, a drawback to the randomized response mechanism is that the estimation error quickly increases with the population size due to the underlying truthful distribution distortion. For example, say we are interested in how many vehicles are at a popular stretch of the highway. Say we configure $\pi_1=0.85$ and $\pi_2=0.3$. We query 10,000 vehicles asking for their current location and only 100 vehicles are at the particular area we are interested in (i.e., 1\% of the population truthfully responds ``Yes"). The standard deviation due to the privacy noise will be 21 \footnote{($\sqrt{(1-0.85) \times 0.3 \times 10,000}$)} which is slightly tolerable. However, a query over one million vehicles (now only 0.01\% of the population truthfully responds ``Yes") will incur a standard deviation of 212. The estimate of the ground truth (100) will incur a \textit{large} absolute error when the aggregated privatized responses are two or even three standard deviations (i.e., 95\% or 99\% of the time) away from the expected value, as the mechanism subtracts \textit{only} the expected value of the noise.

We desire better calibration over the privacy mechanism and a mechanism which maintains constant error as the $No_{pop}$ population scales up. We  introduce the \titlename mechanism in the next section. Though first, we examine how to ensure that data owners are able to privately upload their responses.

\subsection{Private Upload}
\label{sec:privateupload}

\begin{figure}[t!]
    \includegraphics[width=1\columnwidth]{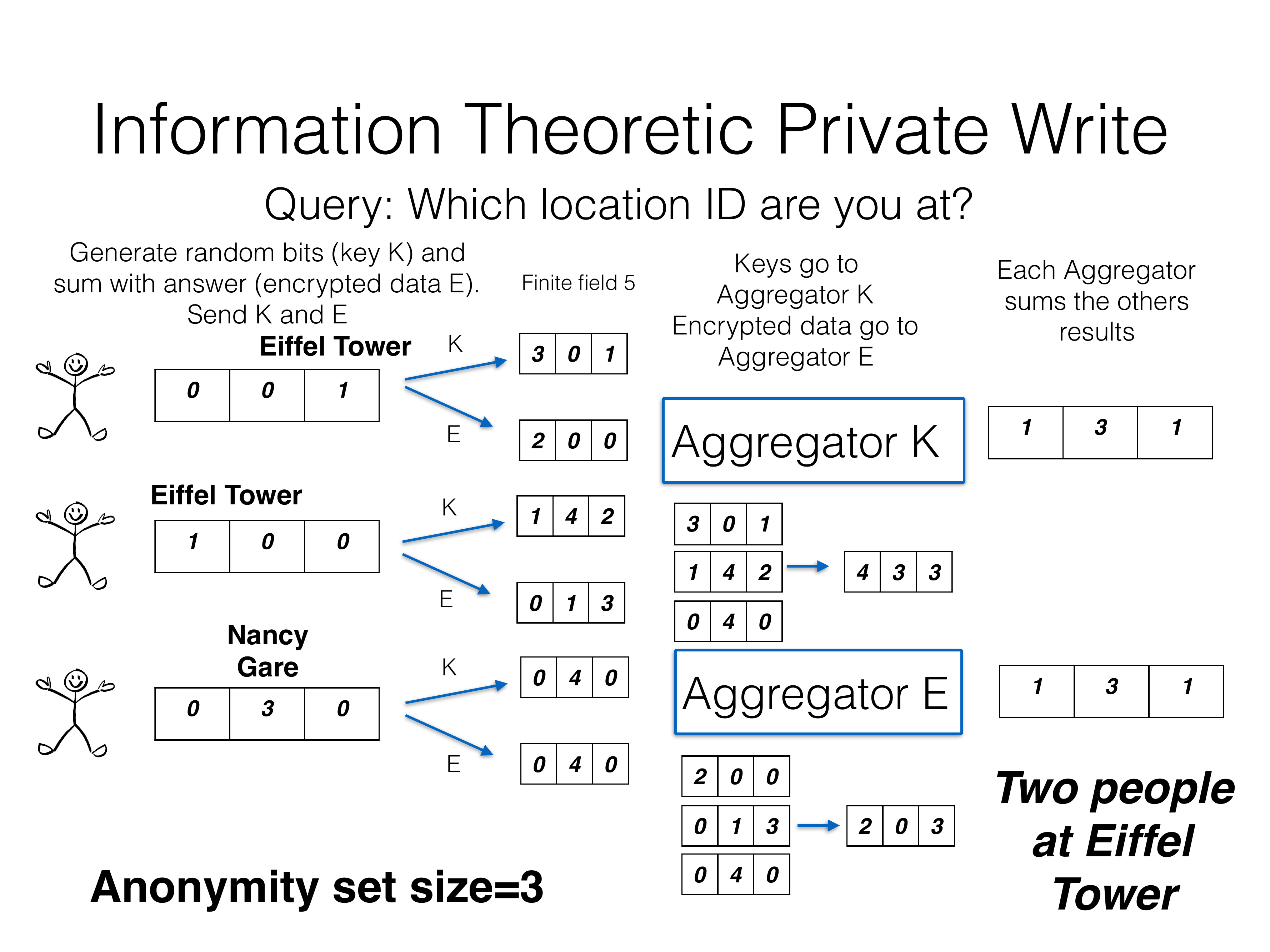}
    \caption{\textbf{Each data owner uniformly at random selects a slot to write their location ID. The aggregators are unable to determine which data owner wrote to a particular slot, as long as there is one honest aggregator who does not collude. The aggregate count of each location ID is computed as the final step.}}
        \label{fig:info-theoretic}
\end{figure}

Now consider that we would like to write into a database without any of the database operators learning which row we wrote into. Knowledge of the row and thus data uploaded by the owner would allow the operator to track the owner over subsequent epochs even if the data is privatized.

We assume a distributed database setting such that, so long as one database operator remains honest and does not collude with other database operators, it is not possible to learn which database row was written into (as long as there are at least two data owners participating). The data owner should continuously re-select a new database row at random every epoch.

~ \\
\noindent \textbf{(Two Party)} Let us first consider two operators and two databases.  The  protocol proceeds as follows.

Assume the database is represented by $n$ rows. Each data owner uploads a message of size $m$ bits, in a randomly chosen row. Without loss of generality, for our example we describe now we assume $m$ is one bit. Thus the database is a bitstring. Extending to a message size of more than one bit would only require a larger finite field (instead of finite field size 2 we could choose a prime number larger than the desired message size in bits).

Each data owner begins with a bitstring of length $n$ (the size of the database). The data owner uniformly at random selects an index of the bitstring and sets its message value (assume it is to 1). Every other index is set to 0.

Next, the data owner creates a key by generating a random bitstring of length $n$.

The data owner then XORs the randomly generated bitstring key with the bitstring containing the message (that has only one index set) to produce the encrypted bitstring.

The data owner then transmits the encrypted bitstring to one database operator and the key bitstring to the other database operator. The data owner may randomly decide which database operator to send the encrypted and key bitstrings.

The same process repeats for each data owner. That is, a second data owner repeats the process of uniformly at random selecting an index of the bitstring to set to 1, generating a key bitstring, and encrypting the bitstring.

As the database operators receive each bitstring (either encrypted or key bitstring), each database operator cumulatively XORs the bitstrings.

Finally, at the end of an agreed epoch, the database operators share the cumulatively XORed bitstrings with each other. By doing so, they are able to reconstruct a database consisting of each data owners message at their specified indexes. The privacy guarantee is that the database operators are unable to determine which data owner wrote to which index, as long as there are at least two participating data owners and there is at least one database operator that does not collude with any other. (There is also an assumption that the data owners do not write to the same index, though collisions can be probabilistically avoided by sizing the database large enough to minimize the likelihood of collisions).

~ \\
\noindent \textbf{(Multi-party)} Now, to generalize the 2 database operators to Z database operators, the protocol proceeds as follows.

The data owner has a bitstring of length n. The data owner uniformly at random selects the index of the bitstring to write to and sets the value to 1. Every other index is set to 0.

Next, the data owner generates $Z-1$ random bitstrings of length n. These bitstrings serve as the ``virtual" single key.

The data owner cumulatively XORs the $Z-1$ bitstring keys with the bitstring containing the message (where only index is set) to produce the encrypted bitstring.

The data owner transmits each separate bitstring key to a different database operator and the encrypted bitstring to the other database operator. The data owner may randomly decide which database operator to send the encrypted and key bitstrings.

The multi-party protocol then proceeds the same as the previous two database operator case. Each database operator cumulatively XORs the received bitstrings and then shares the cumulative XOR results with each other to reconstruct the database. The privacy guarantee holds that each database operator is unable to determine which data owner wrote to which index, as long as there are at least two participating data owners and there is at least one database operator that does not collude with any other.

~ \\
\noindent \textbf{(Key Size)} The issue is that the key size is the length of the database n. Suppose the number of data owners is on the order of millions and the database row size is several hundred bits. The bitstring size will be of the order of several hundred MBs, which is prohibitively expensive for mobile devices and edge networks continually uploading every few minutes.

We could compress each of the key bitstrings by using a pseudorandom generator (PRG) for the Z-1 keys. However, we somehow must compress the bitstring containing the message (that has only one index set). Unfortunately, by definition of a PRG, it is computationally difficult to generate a PRG seed that expands to the desired bitstring. We must utilize a more sophisticated approached to enable cryptographic compression of the bitstring. We utilize a new primitive named Function Secret Sharing (FSS)~\cite{DBLP:conf/eurocrypt/BoyleGI15} and show a performance optimization by selective choice of the parameters in Section~\ref{sec:privatedataupload}.
 \section{\titlename Mechanism}
\label{sec:privacymechanism}

We now describe our \titlename mechanism that achieves constant error even where the population which does not truthfully respond ``Yes" ($No_{pop}$) increases. We illustrate the scheme using location coordinate data, although the scheme can be employed for all real valued data.

~\\ \noindent \textbf{Illustration.} To illustrate and demonstrate the \titlename mechanism, we employ the following example. Suppose we are interested in the distribution of vehicles across London. We query every vehicle in London asking for their location coordinates. Each data owner responds to a binary version of the binary query such as ``Are you at the London Bridge?". The mechanism has two rounds and proceeds as follows.

Suppose a particular data owner is at London Bridge. First, the location is discretized to a location identifier (ID) as described in Section~\ref{sec:discretization}. In this case the location ID is 8 as shown in Figure~\ref{fig:discretization}.

Next, the data owner tosses a multi-sided die. One side samples whether the data owner should respond truthfully for their location ID. The remaining sides selects a location ID for the data owner to respond.

Suppose in the first round the data owner is sampled and selected. The data owner should respond ``Yes" (they are at London Bridge).

In the second round the sampled data owner should abstain from responding at all.

A privatized sum is computed by aggregating the ``Yes" counts in each round.

Finally, estimation occurs by subtracting the privatized sum in round one from round two and dividing by the sampling parameter.

The following three privacy observations are made. First, a majority of the population provides privacy noise by randomly responding either ``Yes" or ``No" regardless of their truthful response. Second, plausible deniability is provided as each data owner probabilistically responds opposite of their truthful response. Finally, \textit{every} data owner acts as a potential candidate for the truthful population. Our assumption is that every data owner is active in both rounds and only the aggregate counts are released.

\subsection{Binary Value}
\label{sec:binaryvalue}

We now formally introduce the binary value \titlename mechanism whereby a data owner responds either ``No" or ``Yes", either 0 or 1 respectively.

~ \\
\noindent \textbf{(Round One)} In the first round each data owner tosses a three sided die with probabilities $\pi_{s}$, $\pi_{Yes}$, and $\pi_{No}$. Let $\pi_{s}$ be the probability that a data owner truthfully responds. Otherwise, regardless of their truthful response let $\pi_{Yes}$ be the probability that a data owner randomly responds ``Yes" and $\pi_{No}$ be the probability that a data owner randomly responds ``No".

\begin{equation}
  Round~One_{Yes} =
  \begin{cases}
    \textbf{1} & \textbf{with probability $\pi_{s}$} \\
    1 & \text{with probability $\pi_{Yes}$} \\
    0 & \text{with probability $\pi_{No}$}
\end{cases}
\end{equation}

\begin{equation}
  Round~One_{No} =
  \begin{cases}
    1 & \text{with probability $\pi_{Yes}$} \\
    \textbf{0} & \textbf{with probability $\pi_{s}$} \\
    0 & \text{with probability $\pi_{No}$}
\end{cases}
\end{equation}

At this point, privacy noise has been added and thus the underlying truthful distribution is becoming distorted as the number of non-truthful data owners participate. The distortion makes it difficult to estimate the the underlying truthful distribution as we have one equation and two variables (number of truthful and non-truthful data owners).

Thus, we execute a second round while fixing the two variables enabling us to solve for the truthful population estimate.

~ \\
\noindent \textbf{(Round Two)} In the second round only the data owner which was selected and sampled with probability $\pi_{s}$ does not participate (effectively writes 0). The remaining data owners stay with the responses from round one.

\begin{equation}
  Round~Two =
  \begin{cases}
    \mathbf{\varnothing} & \textbf{with probability $\pi_{s}$} \\
    1 & \text{with probability $\pi_{Yes}$} \\
    0 & \text{with probability $\pi_{No}$}
\end{cases}
\end{equation}

Now combining the second round with the first round we obtain accurate estimations as we see below.

~ \\
\noindent \textbf{(Expected Values)} We now formulate the expected values as follows.  The subscript refers to the round number. That is, $1_{1}$ refers to output 1 and round 1. The first round of expected values are:

\begin{align}
\begin{split}
\label{eq:roundonesinglequery}
E[1_1] & = \pi_{Yes} \times \mathit{TOTAL_{pop}} + \pi_{s} \times \mathit{Yes_{pop}}  \\
E[0_1] & = \pi_{No} \times \mathit{TOTAL_{pop}} + \pi_{s} \times \mathit{No_{pop}}
\end{split}
\end{align}

That is, both the $Yes_{pop}$ and $No_{pop}$ contribute both ``Yes" and ``No" responses while a small subpopulation responds truthfully.

The second round of expected values are:

\begin{align}
\begin{split}
\label{eq:roundtwosinglequery}
E[1_2] & = \pi_{Yes} \times \mathit{TOTAL_{pop}} \\
E[0_2] & = \pi_{No} \times \mathit{TOTAL_{pop}}
\end{split}
\end{align}

That is, the small subpopulation from round 1 samples out and does not participate (effectively writes 0). The remaining data owners randomly respond ``Yes" or ``No" while remaining at their round one responses.

~ \\
\noindent \textbf{(Estimator)} We solve for the $Yes_{pop}$ population by subtracting round one by round two as follows.  Let $Private~Sum_{``Yes",1}$ refer to the aggregated privatized counts for output space ``Yes" and round 1.

\begin{align}
\begin{split}
Yes_{pop} & = \frac{Private~Sum_{``Yes",1} - Private~Sum_{``Yes",2}}{\pi_{s}}
\end{split}
\end{align}

That is, we subtract the privatized sum of output space ``Yes" round 1 from the output space ``Yes" of round 2. The result is the sampled ``Yes" aggregate. We then obtain the estimation by dividing by the sampling parameter.

The sampling error affects only the $Yes_{pop}$ as seen in Equation~\ref{eq:roundonesinglequery}. Thus we are able to scale the $No_{pop}$ yet retain constant error. Plausible deniability is provided as each data owner may respond to either output space based on the coin toss parameters.

\subsection{Multiple (Simultaneous) Values}

We now examine how to privatize the multiple choice scenario whereby there are multiple values and the data owner should select a single value. We extend the binary value mechanism defined in the previous section. Multiple values are applicable to most real-world scenarios (as opposed to the binary value mechanism). The location coordinate grid scenario, explained in Section~\ref{sec:discretization} and illustrated in Figure~\ref{fig:discretization}, explains a scenario where there are multiple locations (i.e., location IDs) and the data owner is currently at a single location ID. Recall that the data owner's truthful response is discretized to an integer value greater than 0.

However, we desire more than simply randomizing multiple choices. The \titlename mechanism has each data owner respond with \textit{multiple, simultaneous, and contradictory} responses while maintaining constant error. For example, if there are 9 locations, each data owner probabilistically responds they may be in 9 locations simultaneously.

Suppose a particular data owner is at the London Bridge. The first round proceeds as follows. For the location they are currently at (e.g., London Bridge) the data owner flips a biased coin and if heads responds truthfully. When queried about other locations the data owner randomly also responds they are at the location.

Say the particular data owner  from Figure~\ref{fig:discretization} was sampled and selected. Their response should be with location ID 8. In addition, say they were selected for location IDs 1,2,4,5. Thus, the round one response would be 1,2,4,5,8.

In the second round, the data owner should \textit{not} respond they are at the London Bridge. The remaining responses stay. Thus, the round two response would be 1,2,4,5.

A privatized sum is computed by aggregating the location ID counts in each round.

The estimated count for each location ID value is then calculated by subtracting the privatized sum of the second round from the first round and then dividing by the sampling parameter.

There are several privacy observations. Similar to the binary value scenario in Section~\ref{sec:binaryvalue}, both privacy noise and plausible deniability is provided. However, now a data owner responds with multiple \textit{contradictory} responses claiming to be in multiple locations at once. At the same time in the second round, \textit{all} selected and sampled data owners across every location will silently not participate. Every data owner now blends with every other data owner.

We now formally describe the multiple (simultaneous) values \titlename mechanism. 

~ \\
\noindent \textbf{(Round One)} Let $V$ represent all outputs for which the data owner does not truthfully respond ``Yes". Let $V'$ be the special output for which the data owner truthfully responds ``Yes".

In the first round the data owner should truthfully respond according to the sampling parameter only for their truthful output value. For all other values the data owner randomly responds regardless of their truthful value.

\begin{equation}
  Round~One =
  \begin{cases}
    V' & \text{with probability $\pi_{s}$} \\
    V, V' & \text{with probability $\pi_{V}$} \\
    0 & \text{with probability $1-\pi_{V}-\pi_{s}$}
\end{cases}
\end{equation}

That is, a data owner responds with multiple contradictory responses.

~ \\
\noindent \textbf{(Round Two)} In the second round all data owners stay with their round one response and respond randomly regardless of their underlying truthful response. All the data owners that were sampled and selected to respond truthfully do not participate in the second round.

\begin{equation}
  Round~Two =
  \begin{cases}
    \mathbf{\varnothing} & \textbf{with probability $\pi_{s}$} \\
    V,V' & \text{with probability $\pi_{V}$} \\
    0 & \text{with probability $1-\pi_{V}-\pi_{s}$}
  \end{cases}
\end{equation}

That is, all truthful responses equally fall out from the equation.

~ \\
\noindent \textbf{(Expected Values)}  The first round of expected values are as follows.

\begin{align}
\begin{split}
E[V_{1}] & = \pi_{V} \times \mathit{TOTAL} + \pi_{s} \times \mathit{Yes_{pop}} \\
\end{split}
\end{align}

That is, for each value both populations randomly contribute, with a small percentage contributing by responding truthfully. 

The second round of expected values are all the same regardless of the underlying truthful response:

\begin{align}
\begin{split}
E[V_{2}] & = \pi_{V} \times \mathit{TOTAL}
\end{split}
\end{align}

That is, everyone randomly contributes. The sampled and selected percentage that truthfully responded in round one do not participate (effectively write 0) and respond now in round two.

~ \\
\noindent \textbf{(Estimator)} To solve for the \textit{YES} population we subtract the second round from the first round and iterate for each output value as follows:

\begin{align}
\begin{split}
\mathit{YES} & = \frac{Private~Sum_{``Yes",1} - Private~Sum_{``Yes",2}}{\pi_{s}}
\end{split}
\end{align}

The sampled and selected population, by not participating in round two (effectively write 0), allows us to baseline the privacy noise and perform estimation for the sampled truthful population.

\subsection{Differential Privacy Guarantee}
\label{sec:dpprivacyguarantee}

\titlename satisfies differential privacy as we show in this section. We first examine the binary value mechanism and then the multiple value mechanism. The definition of differential privacy can be found in the Appendix~\ref{sec:differentialprivacy}.

~ \\
\noindent \textbf{(Binary Value)} The differential privacy leakage is measured as the maximum ratio of the binary output space given the underlying truthful answer is ``Yes" and ``No" respectively.

In round one, the output space ``Yes" is slightly more likely as the truthful response is sampled in addition to being responded randomly. In round two, there is no privacy leakage as both output space ``Yes" and ``No" are both equally likely and indistinguishable given the truthful answer is either ``Yes" or ``No" respectively. 

Thus, we are interested in the privacy leakage of output 1 round 1  ($1_{1}$) as follows:

\begin{equation}
\epsilon_{DP} = \max \Bigg({\ln \bigg(\frac{\Pr[1_{1} | ``Yes"]} {\Pr[1_{1} | ``No"]} \bigg),\ln \bigg(\frac{\Pr[1_{1} | ``No"]} {\Pr[1_{1} | ``Yes"]} \bigg)} \Bigg)
\end{equation}

\begin{equation}
\frac{\Pr[1_{1} | ``Yes"]} {\Pr[1_{1} | ``No"]} = \frac{ \pi_{V'} + \pi_{s}}{ \pi_{V'} }
\end{equation}

\begin{equation}
\frac{\Pr[1_{1} | ``No"]} {\Pr[1_{1} | ``Yes"]}  = \frac{ \pi_{V'} }{ \pi_{V'} + \pi_{s} }
\end{equation}

\begin{equation}
\label{eq:binaryleakage}
\epsilon_{DP} = \max \Bigg( \ln \bigg(\frac{ \pi_{V'} + \pi_{s}}{ \pi_{V'}}\bigg),\ln \bigg(\frac{ \pi_{V'} }{ \pi_{V'} + \pi_{s} }\bigg) \Bigg) 
\end{equation}

~ \\
\noindent \textbf{(Multiple (Simultaneous) Values)} The differential privacy leakage is measured as the maximum ratio of the multiple output space given the underlying truthful answer is any combination of two values of the output space of size $V$.

In round one, for any two outputs whereby a data owner would not truthfully respond with those values, the outputs are indistinguishable and there is no privacy leakage as the response is chosen randomly. The only privacy leakage occurs when the data owner truthfully responds for their output space.

For \textbf{output V', round 1 ($\mathbf{V'_{1}}$)}, the privacy leakage is as follows:

\begin{equation}
\epsilon_{DP} = \max \Bigg({\ln \bigg(\frac{\Pr[V'_{1} | V']} {\Pr[V'_{1} | \neg V']} \bigg),\ln \bigg(\frac{\Pr[V'_{1} | \neg V']} {\Pr[V'_{1} | V']} \bigg)} \Bigg)
\end{equation}

\begin{equation}
\frac{\Pr[V'_{1} | V']} {\Pr[V'_{1} | \neg V']} = \frac{ \pi_{V'} + \pi_{s}}{ \pi_{V'} }
\end{equation}

\begin{equation}
\frac{\Pr[V'_{1} | \neg V']} {\Pr[V'_{1} | V']}  = \frac{ \pi_{V'} }{ \pi_{V'} + \pi_{s} }
\end{equation}

\begin{equation}
\label{eq:leakagemultipleround1}
\epsilon_{DP} = \max \Bigg( \ln \bigg(\frac{ \pi_{V'} + \pi_{s}}{ \pi_{V'} }\bigg),\ln \bigg(\frac{ \pi_{V'} }{ \pi_{V'} + \pi_{s} }\bigg) \Bigg) 
\end{equation}

In round two, again there is no privacy leakage for those values the data owner would not truthfully respond as the response is chosen randomly. The only privacy leakage occurs for those sampled and selected data owners that do not participate (effectively write 0).

For \textbf{output V', round 2 ($\mathbf{V'_{2}}$)}, the privacy leakage is as follows:

\begin{equation}
\epsilon_{DP} = \max \Bigg({\ln \bigg(\frac{\Pr[V'_{2} | V']} {\Pr[V'_{2} | \neg V']} \bigg),\ln \bigg(\frac{\Pr[V'_{2} | \neg V']} {\Pr[V'_{2} | V']} \bigg)} \Bigg)
\end{equation}

\begin{equation}
\frac{\Pr[V'_{2} | V']} {\Pr[V'_{2} | \neg V']} = \frac{ \pi_{V'} - \pi_{s}}{ \pi_{V'} }
\end{equation}

\begin{equation}
\frac{\Pr[V'_{2} | \neg V']} {\Pr[V'_{2} | V']}  = \frac{ \pi_{V'} }{ \pi_{V'} - \pi_{s} }
\end{equation}

\begin{equation}
\label{eq:leakagemultipleround2}
\epsilon_{DP} = \max \Bigg( \ln \bigg(\frac{ \pi_{V'} - \pi_{s}}{ \pi_{V'} }\bigg),\ln \bigg(\frac{ \pi_{V'} }{ \pi_{V'} - \pi_{s} }\bigg) \Bigg) 
\end{equation}

We then take the maximum leakage of round 1 (Equation~\ref{eq:leakagemultipleround1}) and round 2 (Equation~\ref{eq:leakagemultipleround2}).

 \subsection{Calibrating Noise to Population Size}

We now examine how to add privacy noise to the second round independent of the first round. Our query expansion will continue to add the ``No" population which means that we must calibrate the sampling rate to avoid incurring a large standard error due to variance.

Thus, we calibrate the sampling rate standard deviation to the expected population size. This means that when the query expansion is being performed, some effort must be made to estimate the target population size beforehand. However, in cases a meaningful estimate is not able to be performed, the issuer of the query will need to issue a probe query and then issue the calibrated query.

~ \\
\noindent \textbf{(Round One)} In the first round each data owner tosses a three sided die with a sampling probability $\pi_{s}$, $\pi_{Yes}$, and $\pi_{No}$. Let $\pi_{s}$ be the sampling probability that the data owner is forced to respond ``Yes". Let $\pi_{No}$ be the probability that a data owner randomly responds ``No". We are careful to calibrate the sampling rate to reduce the standard deviation according to the expected population size due to the query expansion and expected subpopulations (Yes and No).

\begin{equation}
  Round~One_{Yes} =
  \begin{cases}
    \textbf{1} & \textbf{with probability $\pi_{s_{Yes_1}}$} \\
    0 & \text{with probability $\pi_{No}$}
\end{cases}
\end{equation}

\begin{equation}
  Round~One_{No} =
  \begin{cases}
    \textbf{1} & \textbf{with probability $\pi_{s_{No_1}}$} \\  
    0 & \text{with probability $\pi_{No}$}
\end{cases}
\end{equation}

At this point, privacy noise has been added and thus the underlying truthful distribution is becoming distorted as the number of non-truthful data owners participate. The distortion makes it difficult to estimate the the underlying truthful distribution as we have one equation and two variables (number of truthful and non-truthful data owners).

Thus, we execute a second round. We conduct a fresh sample again being careful to  calibrate the sampling rate to minimize the variance for each population enabling us to solve for the truthful population estimate.

~ \\
\noindent \textbf{(Round Two)} In the second round only the data owner which was selected and sampled with probability $\pi_{s}$ does not participate (effectively writes 0). The remaining data owners stay with the responses from round one.

\begin{equation}
  Round~Two_{Yes} =
  \begin{cases}
    \textbf{1} & \textbf{with probability $\pi_{s_{Yes_2}}$} \\
    0 & \text{with probability $\pi_{No}$}
\end{cases}
\end{equation}

\begin{equation}
  Round~Two_{No} =
  \begin{cases}
    \textbf{1} & \textbf{with probability $\pi_{s_{No_2}}$} \\  
    0 & \text{with probability $\pi_{No}$}
\end{cases}
\end{equation}

Now combining the second round with the first round we obtain accurate estimations as we see below.

~ \\
\noindent \textbf{(Expected Values)} We now formulate the expected values as follows.  The subscript refers to the round number. That is, $1_{1}$ refers to output 1 and round 1. The first round of expected values are:

\begin{align}
\begin{split}
\label{eq:roundonesinglequery}
E[1_1] & = \pi_{s_{Yes_1}} \times \mathit{Yes_{pop}}  +  \pi_{s_{No_1}} \times \mathit{No_{pop}} \\
E[0_1] & = \pi_{No} \times \mathit{TOTAL_{pop}}
\end{split}
\end{align}

That is, both the $Yes_{pop}$ and $No_{pop}$ contribute both ``Yes" and ``No" responses.

The second round of expected values are:

\begin{align}
\begin{split}
\label{eq:roundtwosinglequery}
E[1_2] & = \pi_{s_{Yes_2}} \times \mathit{Yes_{pop}}  +  \pi_{s_{No_2}} \times \mathit{No_{pop}} \\
E[0_2] & = \pi_{No} \times \mathit{TOTAL_{pop}}
\end{split}
\end{align}

That is, both the $Yes_{pop}$ and $No_{pop}$ contribute both ``Yes" and ``No" responses.

~ \\
\noindent \textbf{(Estimator)} We solve for the $Yes_{pop}$ population by subtracting round one by round two as follows.  Let $Private~Sum_{``Yes",1}$ refer to the aggregated privatized counts for output space ``Yes" and round 1.

\begin{align}
\begin{split}
Yes_{pop} & = \frac{Private~Sum_{``Yes",1} - Private~Sum_{``Yes",2}}{\pi_{s}}
\end{split}
\end{align}

That is, we subtract the privatized sum of output space ``Yes" round 1 from the output space ``Yes" of round 2. The result is the sampled ``Yes" aggregate. We then obtain the estimation by dividing by the sampling parameter.

We configure $\pi_{s_{Yes_1}}$ according to the desired privacy and accuracy tradeoff. However, we must configure $ \pi_{s_{Yes_2}}, \pi_{s_{No_2}}, and \pi_{s_{No_2}}$ to minimize the standard deviation as we perform a fresh sample each round.

 \section{Private Data Upload}
\label{sec:privatedataupload}

We now describe the Function Secret Sharing (FSS)~\cite{DBLP:conf/eurocrypt/BoyleGI15} primitive and how it enables a nearly square root reduction of the key size. We then introduce our parameter optimization to achieve more than an \textit{order of magnitude improvement} over the default implementation.

\subsection{Function Secret Sharing Background}

Recall our earlier construction in Section~\ref{sec:privateupload} whereby data owners privately upload data into a distributed database without any of the $Z$ database operators learning into which row a particular data owner wrote (assuming there is at least one honest database operator that does not collude and there are at least two honest data owners). Each data owner specifies their upload data by uniformly at random selecting a row from the database to write their message. This selection of a single row $a$ and writing a message $m$ can be viewed as a point function $F_{a}$ whereby $F_{a}(x)=m$ iff $x==a$ and $0$ otherwise.

The key idea of FSS to obtain the key size reduction is the use of a pseudorandom generator (PRG) to compress the key size. However, as we previously saw simply using a PRG alone is not enough as it's not computationally feasible to find $Z$ random seeds that cumulatively XORed together produce a desired output. We can find $Z-1$ random seeds, cumulatively XOR them together, then XOR with the desired output to find the correction word bitstring needed to XOR with the seeds to produce a desired output. It is this observation that multi-party FSS exploits to achieve the nearly square root key size reduction.

FSS addresses the issue of the correction word being the length of the database (thus the key size the length of the database) by the use of a special matrix. Indexing into the matrix is done by using the lower and higher bits of the input to lookup into the matrix. Each matrix row contains a set of PRG seeds. The expansion of a particular subset of the PRG seeds are then combined by XOR with the correction word. The resultant bitstring is then the length of the correction word and contains both the desired output as well as other random noise. Using the higher order bits of $a$ as a lookup into the resultant bitstring will locate the desired output within the resultant bitstring. All other inputs will produce random noise.

Thus, the cryptographic compression is achieved by the using of the PRG combined with correction words with length roughly the square root of the size of the database. Further details can be found in the paper by Boyle et al~\cite{DBLP:conf/eurocrypt/BoyleGI15}.

However, it turns out that by adjusting the length of the correction words and number of PRG seeds greatly impacts the performance of the FSS protocols. We now explain our FSS parameter optimization.

\subsection{Function Secret Sharing Optimization}
\label{sec:fssoptimization}

FSS relies on symmetric cryptography. Thus, we utilize AES in counter mode for the pseudorandom generator.

There are two symmetric cryptography overheads that FSS incurs. The first is that the default FSS evaluation algorithm repeatedly evaluates the same seeds multiple times. The only difference between each evaluation is that different positions of the seed expansion evaluation is used based on the input's lower bits. The second overhead is balancing the number of seeds with seed expansions.

Our algorithm optimization is described in Algorithm~\ref{alg:evaluatesharefss} and Algorithm~\ref{alg:evalfss}. The share generation algorithm~\ref{alg:genfss}, is the same as described in ~\cite{DBLP:conf/eurocrypt/BoyleGI15}. The difference is in the share evaluation. The default implementation performs $2^n$ PRG seed initializations. However, the full PRG evaluation is the same for each value of $\gamma \prime$. Thus, we need to perform $\nu$ PRG seed initializations instead of repeating the same PRG evaluation. We do \textit{one} evaluate per $\delta$, which means that we can reuse the same evaluated output and just take different partitions for varying $\gamma$. The default FSS version evaluates $\delta$ times the same seeds in order to extract the differing $\gamma$ sections.

Our second optimization is balancing the number of total seeds generated and evaluated with the prg expansion length of each seed. Increasing the length of the seed reduces the number of total seeds required. It's faster to expand a single side as opposed to multiple expansions of differing seeds. However, the cost is an increase in the length of the key size. Thus, we chose our parameters by performing microbenchmarks to understand the tradeoffs in. 

\begin{algorithm}
\label{alg:genfss}
\caption{Gen$^{p_i}(1^\lambda,x,y)$: Generate Shares}
\begin{algorithmic}[1]
\item Let $G$ : \{0,1\}$^\lambda$ $\longrightarrow$ \{0,1\}$^{m\mu}$ be a PRG
\item Let $\mu$ $\leftarrow$ $\lceil 2^{n/2} \times 2^{{p-1}/2} \rceil$. Let $\nu \leftarrow \lceil 2^n / \mu \rceil$
\item Use the higher and lower bits of the input $x$ as a pair $x=(\gamma \prime,\delta \prime)$, $\gamma \prime \in \lceil \nu \rceil$ $\delta \prime \in \lceil \mu \rceil$
\item Choose $\nu$ arrays $A_1,...,A_{\nu}$ s.t. $A_{\gamma} \in_R O_p$ and $A_{\gamma \prime} \in_R E_p$ for all $\gamma \prime \neq \gamma$
\item Choose 2$^{p-1}$ random strings $cw_1,...,cw_{2^{p-1}} \in 0,1^{m \mu}$ s.t. $\bigoplus_{j=1}^{2^{p-1}} (cw_j \oplus G(s_{\gamma, j})) = e_{\delta} \cdot b$
\item Set $\sigma_{i,\gamma \prime} \leftarrow (s_{\gamma \prime,1} \cdot A_{\gamma \prime}[i,1]) \parallel ... \parallel (s_{\gamma \prime, 2^{p-1}} \cdot A_{\gamma \prime}[i,2^{p-1}])$ for all 1 $\leq i \leq p$, 1 $ \leq \gamma \prime \leq \nu$.
\item Set  $\sigma_i = \sigma_{i,1} \parallel ... \parallel ... \sigma_{i,\nu}$ for 1 $\leq i \leq$ p
\item Let $k_i = (\sigma_i \parallel cw_1 \parallel ... \parallel cw_{2^{p-1}})$ for 1 $\leq i \leq$ p
\item Return $(k_i,...,k_p)$
\end{algorithmic}
\end{algorithm}

\begin{algorithm}
\label{alg:evaluatesharefss}
\caption{EvaluateShare$^{p_i}(k_i)$: Evaluate Share}
\begin{algorithmic}[1]
\item Let $\mu$ $\leftarrow$ $\lceil 2^{n/2} \times 2^{{p-1}/2} \rceil$. Let $\nu \leftarrow \lceil 2^n / \mu \rceil$
\item Use the higher and lower bits of the input $x$ as a pair $x=(\gamma \prime,\delta \prime)$, $\gamma \prime \in \lceil \nu \rceil$ $\delta \prime \in \lceil \mu \rceil$
\For{$j = 1,...,\nu$}
	\State  $y_j \leftarrow$ Eval($j,k_i$)
	\State Let $result_j \leftarrow (y_j[1] \parallel ... \parallel y_j[\mu])$
\EndFor
\item Return $(result_1,...,result_{\nu})$
\end{algorithmic}
\end{algorithm}

\begin{algorithm}
\label{alg:evalfss}
\caption{Eval$^{p_i} (\nu \prime,k_i)$}
\begin{algorithmic}[1]
\item Let $G$ : \{0,1\}$^\lambda$ $\longrightarrow$ \{0,1\}$^{m\mu}$ be a PRG
\item Parse $k_i$ as $k_i = (\sigma_i, cw_i, ..., cw_{2^{p-1}})$
\item Parse $\sigma_i$ as $\sigma_i = s_{1,1} \parallel ... \parallel s_{1,2^{p-1}} \parallel ... \parallel s_{\nu,2^{p-1}}$
\item Let $y_i \leftarrow \bigoplus_{1 \leq j \leq 2^{p-1}}^{} (cw_j \oplus G(s_{\gamma \prime, j}))$ where $s_{\gamma \prime,j} \neq 0$
\item Return $y_i$
\end{algorithmic}
\end{algorithm}
 \section{Pollution Protection}
\label{sec:pollutionprotection}

In order to guard the private writes against a single data owner writing to multiple rows, we utilize a novel and efficient FSS share verification technique (which does not require any public-key primitives) that is performed by the aggregation parties~\cite{DBLP:conf/ccs/BoyleGI16}. The FSS share verification ensures that each data owner writes a unit vector (i.e., a single row). Details of the MPC technique can be found in the Appendix~\ref{sec:shareverificationexample}. Our evaluation results can be found in Section~\ref{sec:evaluation:scalability}.

Next, to prevent a single answer, such as a large number, from distorting the aggregate sum, we utilize a bit vector response which limits the data owner to only replying ``No'' ('0') or ``Yes'' ('1'). 

Finally, data owners are authenticated to prevent Sybils and multiple responses within a single epoch. The authentication does not allow the aggregators to learn which row a data owner is writing to. Each data owner performs a cryptographic private write that is protected as long as there is at least one honest aggregator who does not collude (as we previously shown in Section~\ref{sec:privatedataupload}).

Defining the error threshold for the number of malicious data owners who falsify their responses (i.e., intentionally answering ``No" instead of ``Yes") is not considered in this work. However, efficient techniques exist which ensure commitment to the randomized response protocol~\cite{Kikuchi:1999:SVP:519623.837364}. \\
 \begin{figure}[!htb]
\centering
    \includegraphics[width=1\columnwidth]{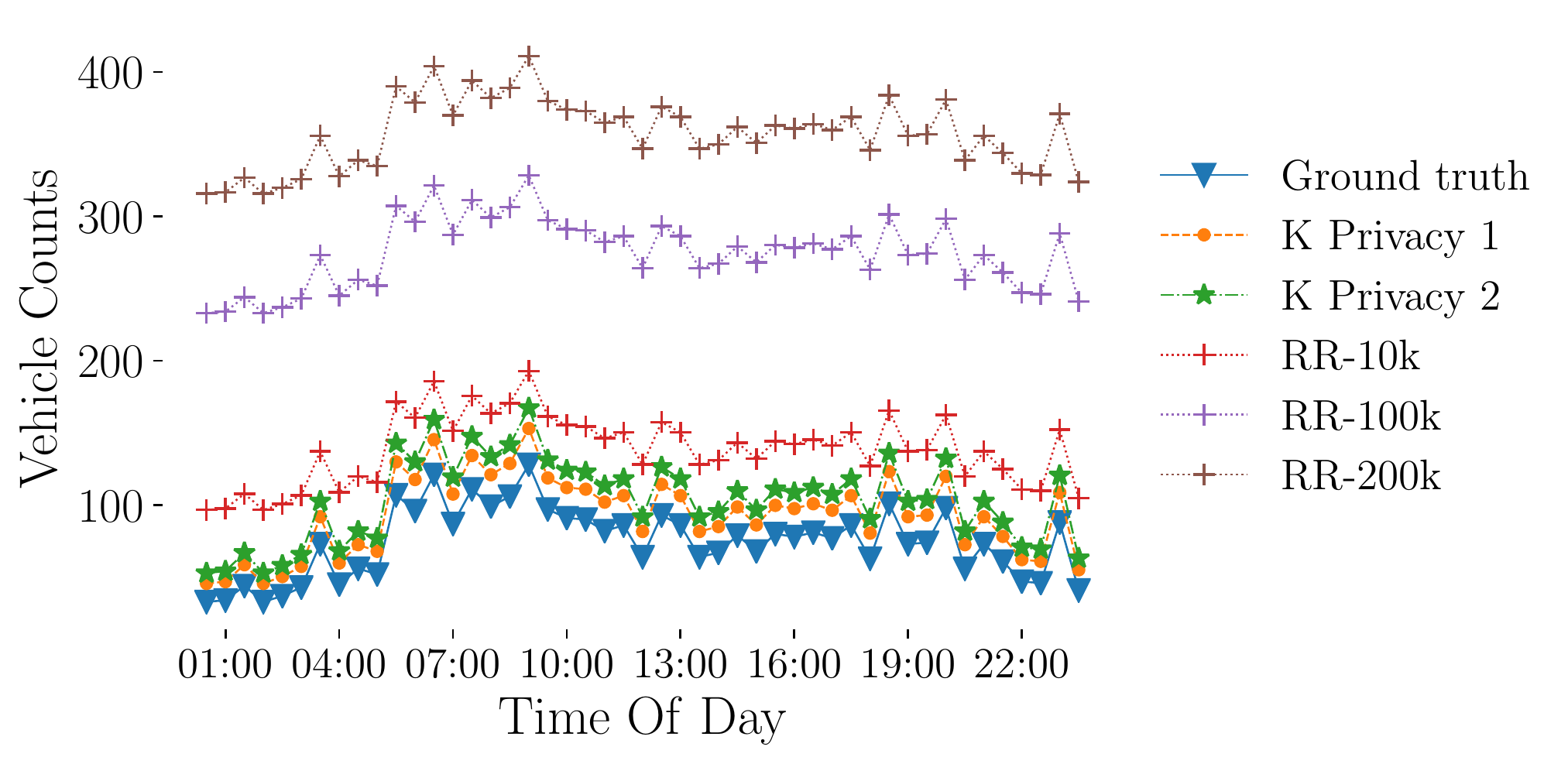}
    \caption{\textbf{(Vehicle counts)} \titlename Each vehicle reports it's current location. }
    \label{fig:pems}
\end{figure}

\begin{figure}[!htb]
\centering
    \includegraphics[width=1\columnwidth]{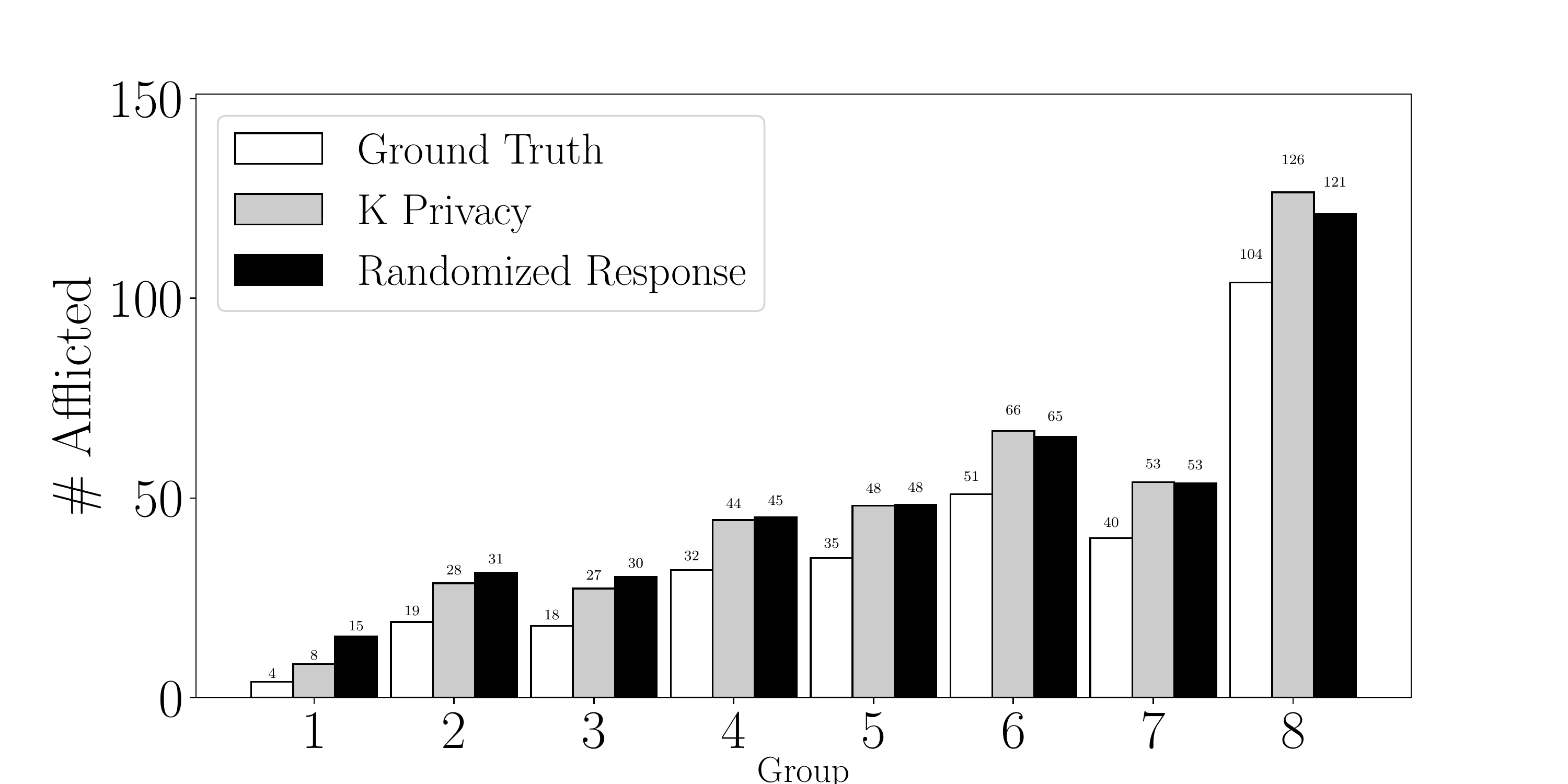}
    \caption{\textbf{(Heart Chest Pain)} Number of individuals out of 303 with specific types of heart related chest pain. }
    \label{fig:heart-chestpain}
\end{figure}

\begin{figure}[!htb]
\centering
    \includegraphics[width=1\columnwidth]{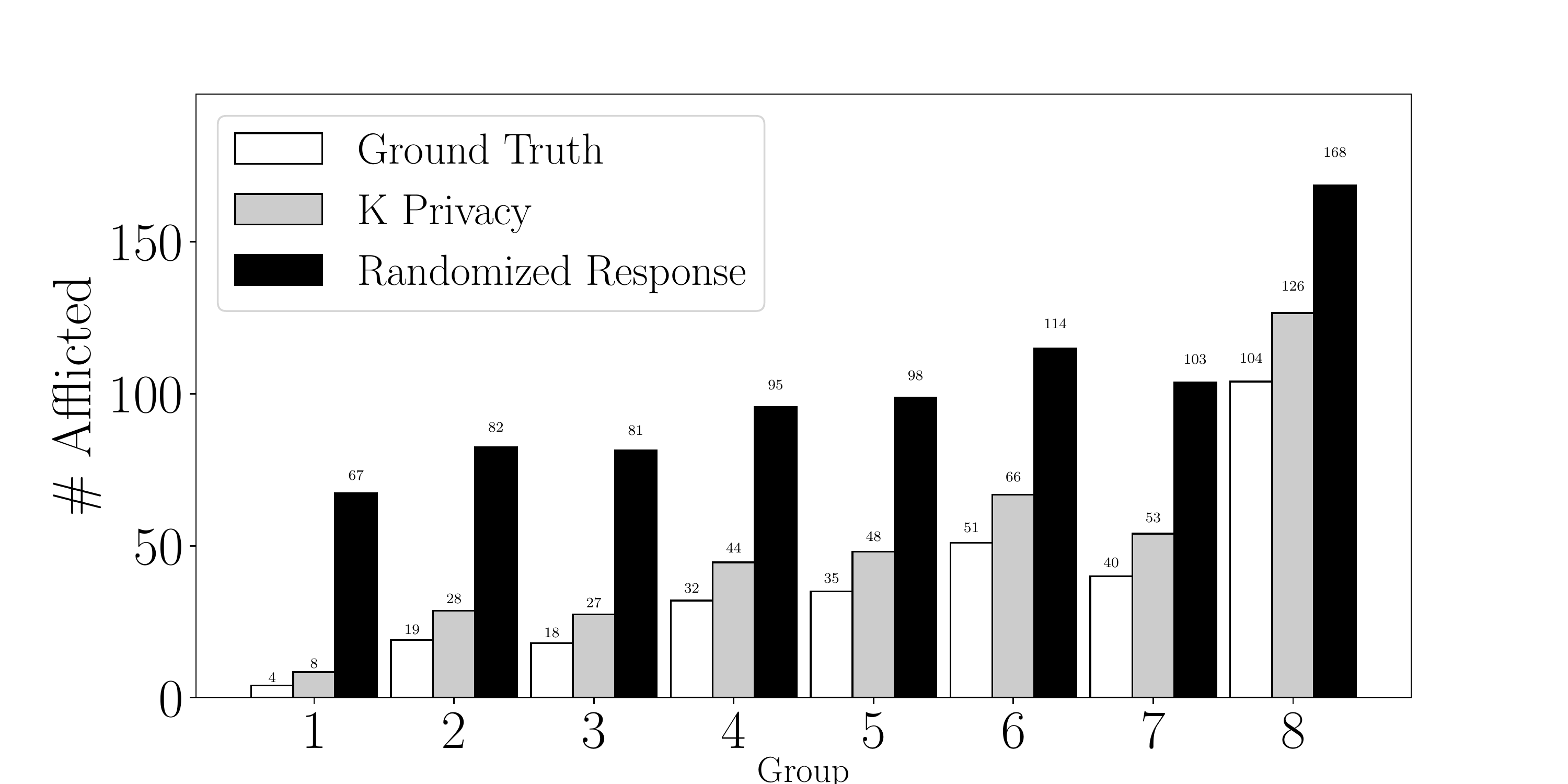}
    \caption{\textbf{(Heart Chest Pain)} Number of individuals out of 10,000 with specific types of heart related chest pain.}
    \label{fig:heart-chestpain-10k}
\end{figure}

\begin{figure}[!htb]
\centering
    \includegraphics[width=1\columnwidth]{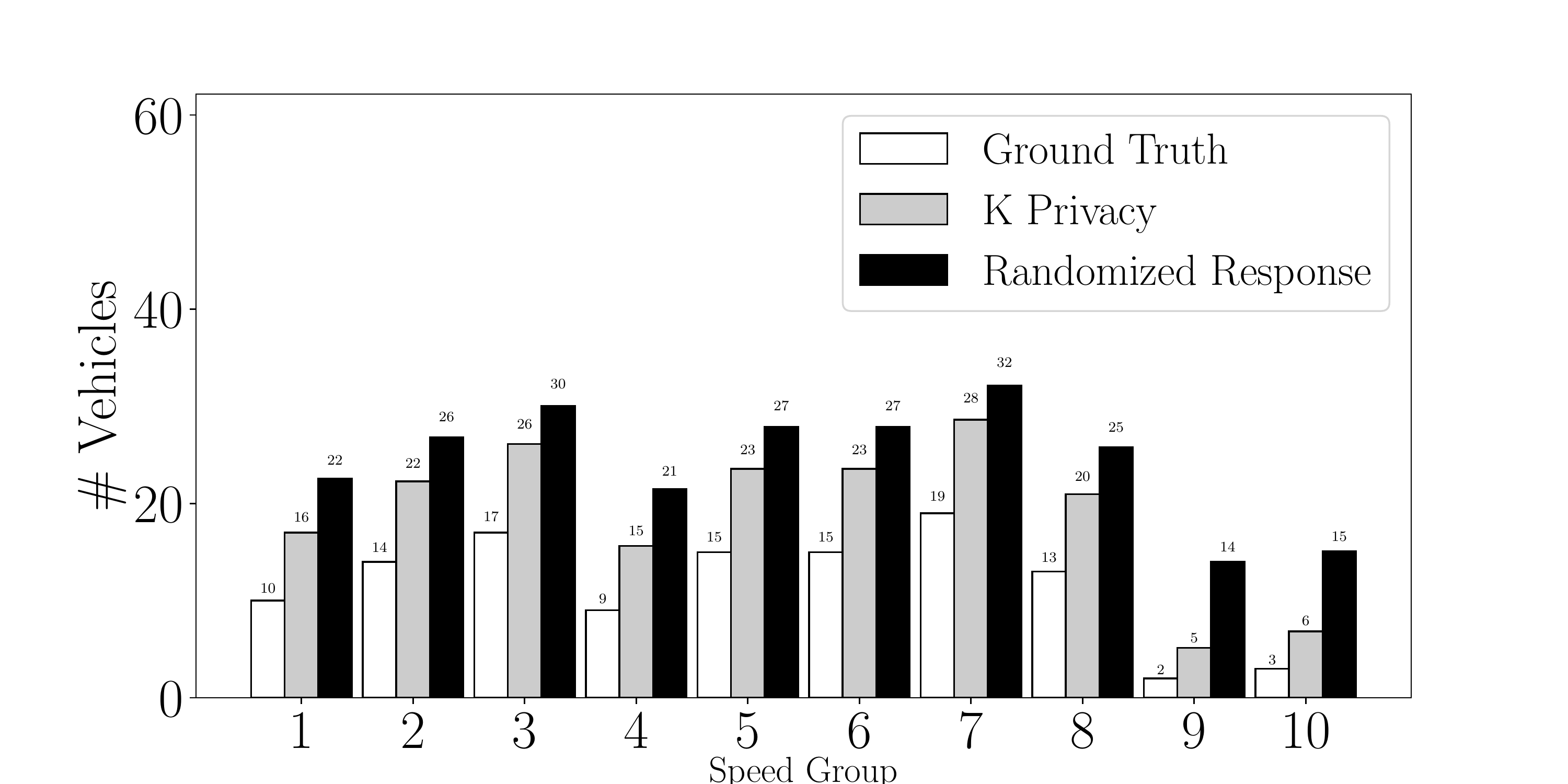}
    \caption{\textbf{(Vehicle Speed Distribution)} Lane 1 speed distribution. }
    \label{fig:lane1-354}
\end{figure}

\begin{figure}[!htb]
\centering
    \includegraphics[width=1\columnwidth]{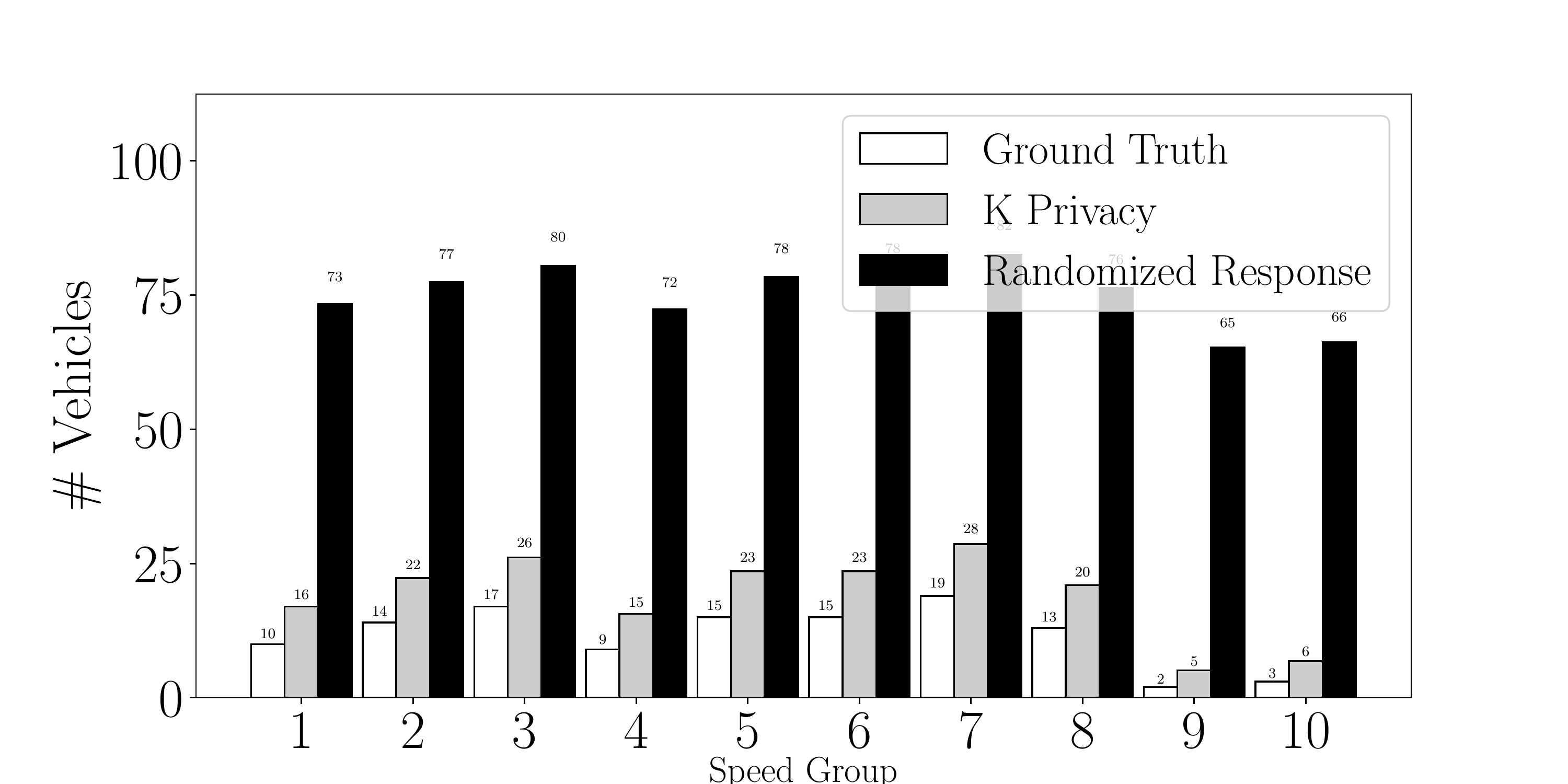}
    \caption{\textbf{(Vehicle Speed Distribution)} Lane 1 speed distribution over 10,000 vehicles (only 354 are currently amongst the queried 3 lanes). }
    \label{fig:lane1-10k}
\end{figure}

\section{Evaluation}
\label{sec:evaluation}

We evaluate the accuracy of the \titlename mechanism. Next, we describe the performance gains of the FSS parameter optimization. Finally, we evaluate the efficiency of the pollution protection technique.

\subsection{Accuracy}
\label{sec:evaluation:accuracy}

~ \\
\noindent \textbf{(PeMS Data)} We evaluate the \titlename mechanism over a real dataset rather than with arbitrary distributions. We utilize the California Transportation Dataset from magnetic pavement sensors\cite{pems} collected in LA$\backslash$Ventura California freeways \cite{cwwpinformation}.  There are a total of 3,865 stations and 999,359 vehicles total. We assign virtual identities to each vehicle. Each vehicle announces the station it is currently at. We select a single popular highway station. Every vehicle at the station reports ``Yes" while every other vehicle in the population truthfully reports ``No". We evaluate over a 24 hour time period. \titlename 1 has a sampling parameter of 45\% and \titlename 2 has a sampling parameter of 25\%. The randomized response mechanism has $\pi_1=0.8$ and $\pi_2=0.2$.

Figure~\ref{fig:pems} compares the \titlename mechanism with the Randomized Response mechanism. \titlename is able to maintain constant error even at 1 million vehicles, while the Randomized Response quickly incurs error. Upper bounds are shown with a 95\% confidence interval.

We next examine the vehicle speed distribution across the freeways at evening rush hour. Figure~\ref{fig:lane1-354} is with the population at the specific stretch of the freeway. Figure~\ref{fig:lane1-10k} expands the query population to 10,000 vehicles (9,646) are not at the particular freeway stretch being monitored. The figures show the speed distribution whereby there are 10 groups for the following speeds $``1-10"$ is group 1, $``11-20"$ is group 2, etc. Upper bounds are shown with a 95\% confidence interval.

~ \\
\noindent \textbf{(Heart Data)} We next evaluate over medical data. We utilize the UCI open data repository~\cite{Lichman:2013} for heart related data. Figure~\ref{fig:heart-chestpain} and Figure~\ref{fig:heart-chestpain-10k} show the number of afflicted data owners with a particular type of chest pain. The four types of chest pain are typical angina, atypical angina, non-anginal pain, and asymptomatic. Each group corresponds to a particular chest pain and gender for a total of eight groups. Figure~\ref{fig:heart-chestpain-10k}  scales the population to 10,000 whereby 303 are the original dataset and the remaining 9,697 data owners provide chaff. The \titlename mechanism maintains constant error and the randomized response quickly incurs error. Upper bounds are shown with a 95\% confidence interval.

\begin{figure}[!htb]
\centering
    \includegraphics[width=1\columnwidth]{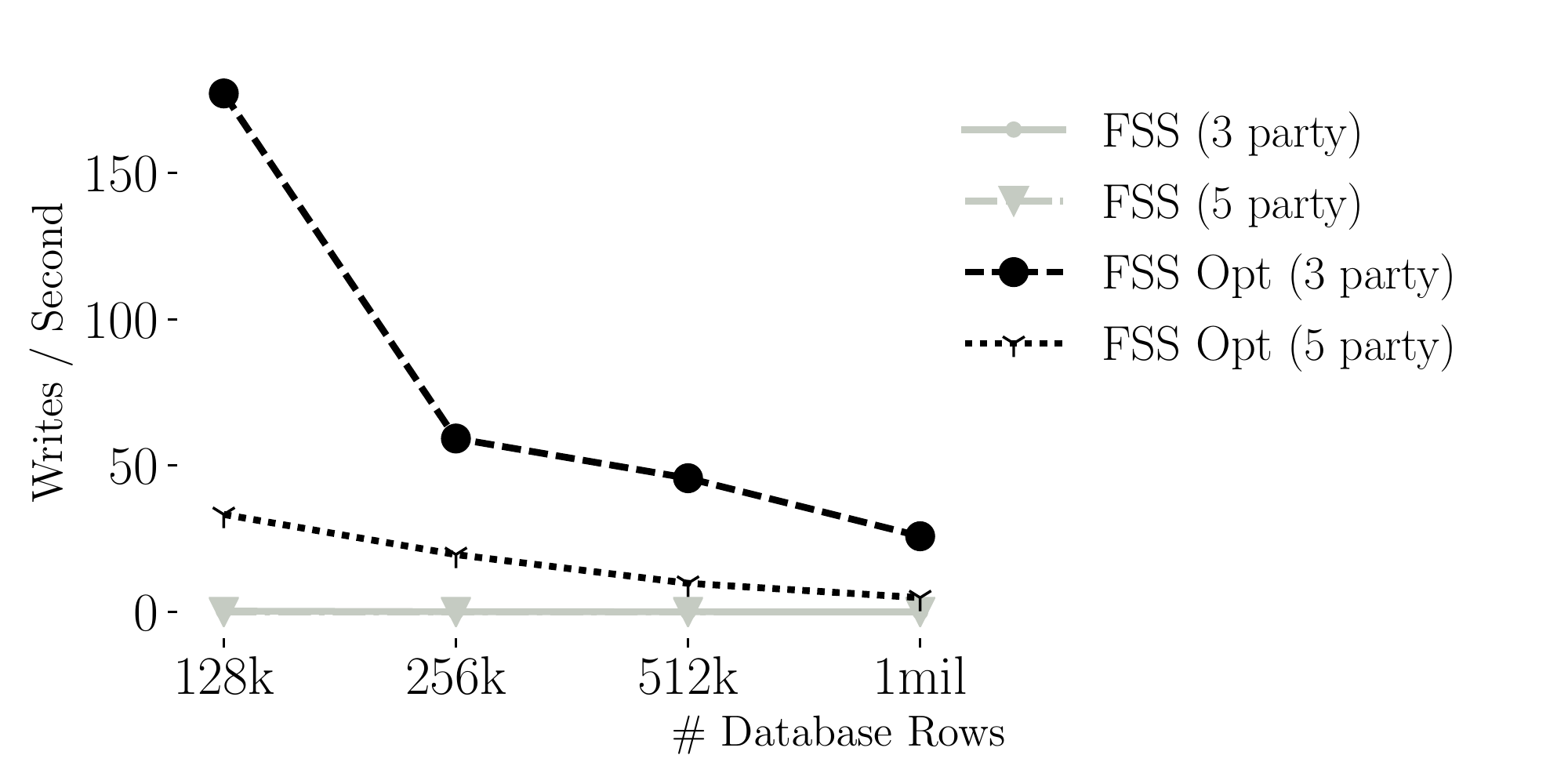}
    \caption{\textbf{(FSS Microbenchmark)} Number of share evaluations (client share uploads) per second. Bigger is better.}
    \label{fig:fss-microbenchmark}
\end{figure}

\begin{figure}[!htb]
\centering
    \includegraphics[width=1\columnwidth]{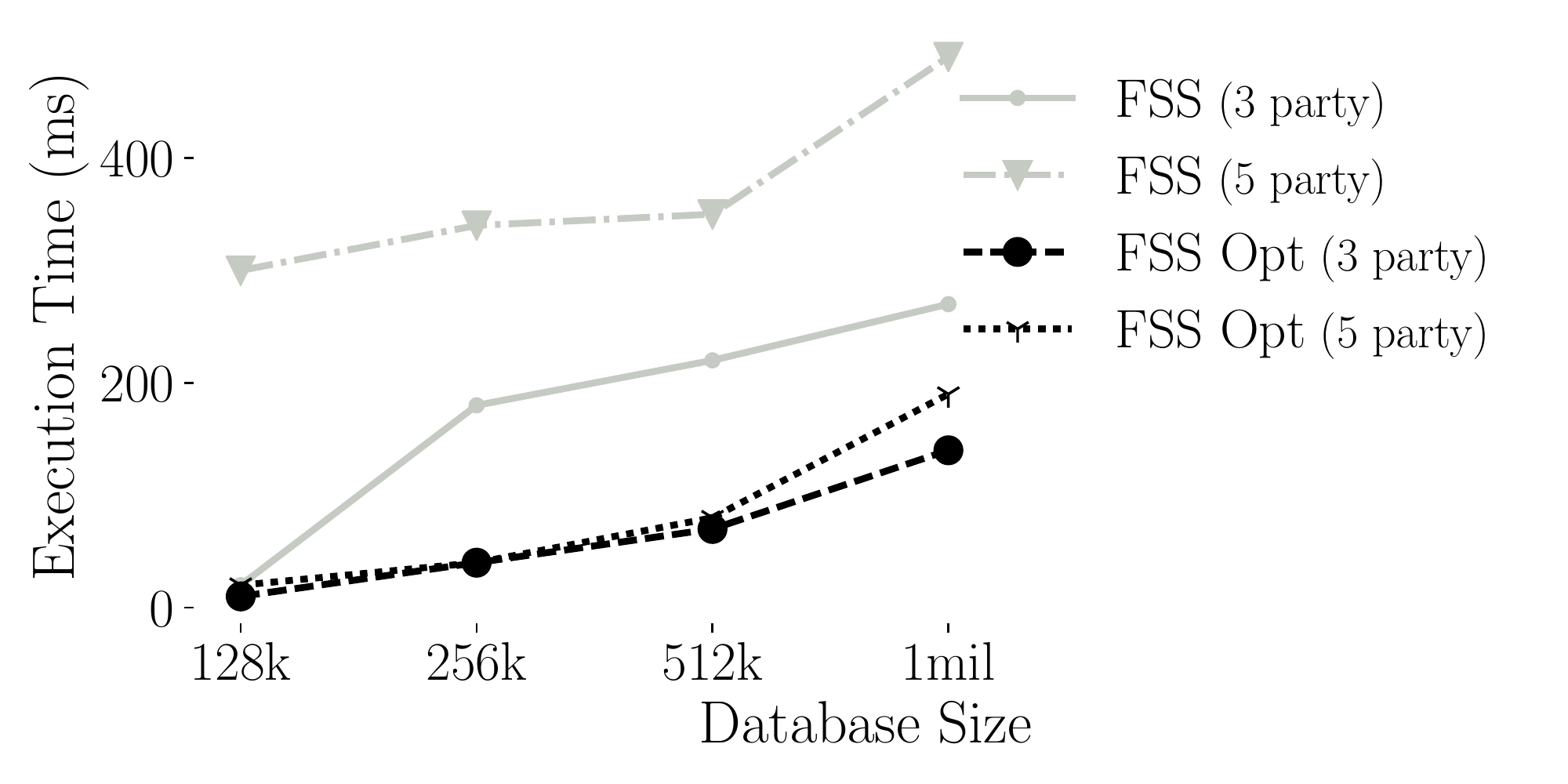}
    \caption{\textbf{FSS Share Generation.}   FSS versus FSS optimized client share upload. Smaller is better. }
    \label{fig:genshares}
\end{figure}

\begin{figure}[!htb]
\centering
    \includegraphics[width=1\columnwidth]{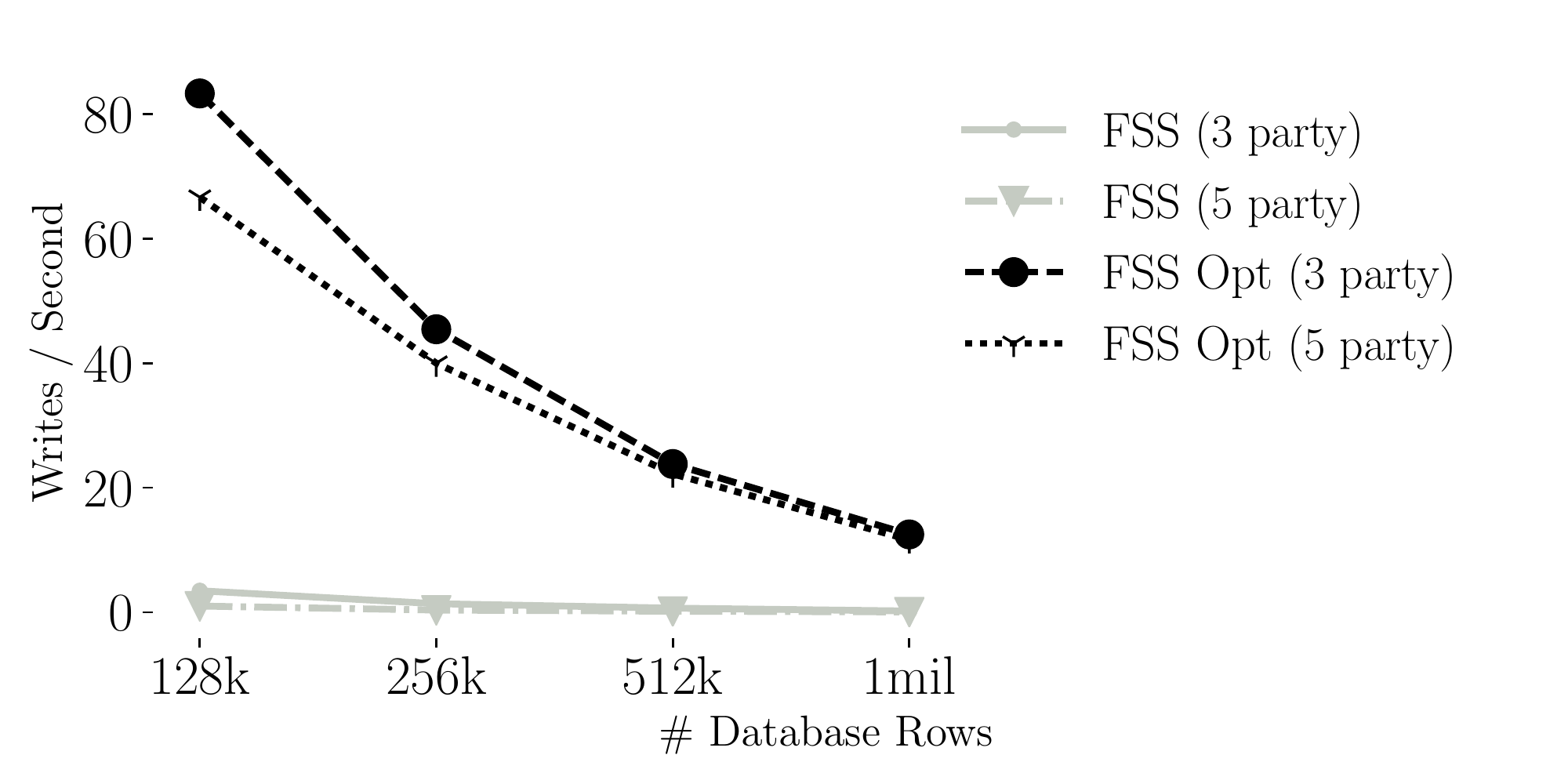}
    \caption{\textbf{FSS Scaling Optimization.}   FSS versus FSS optimized (client share uploads) per second. Bigger is better.}
    \label{fig:fssscaling}
\end{figure}

\begin{figure}[!htb]
\centering
    \includegraphics[width=1\columnwidth]{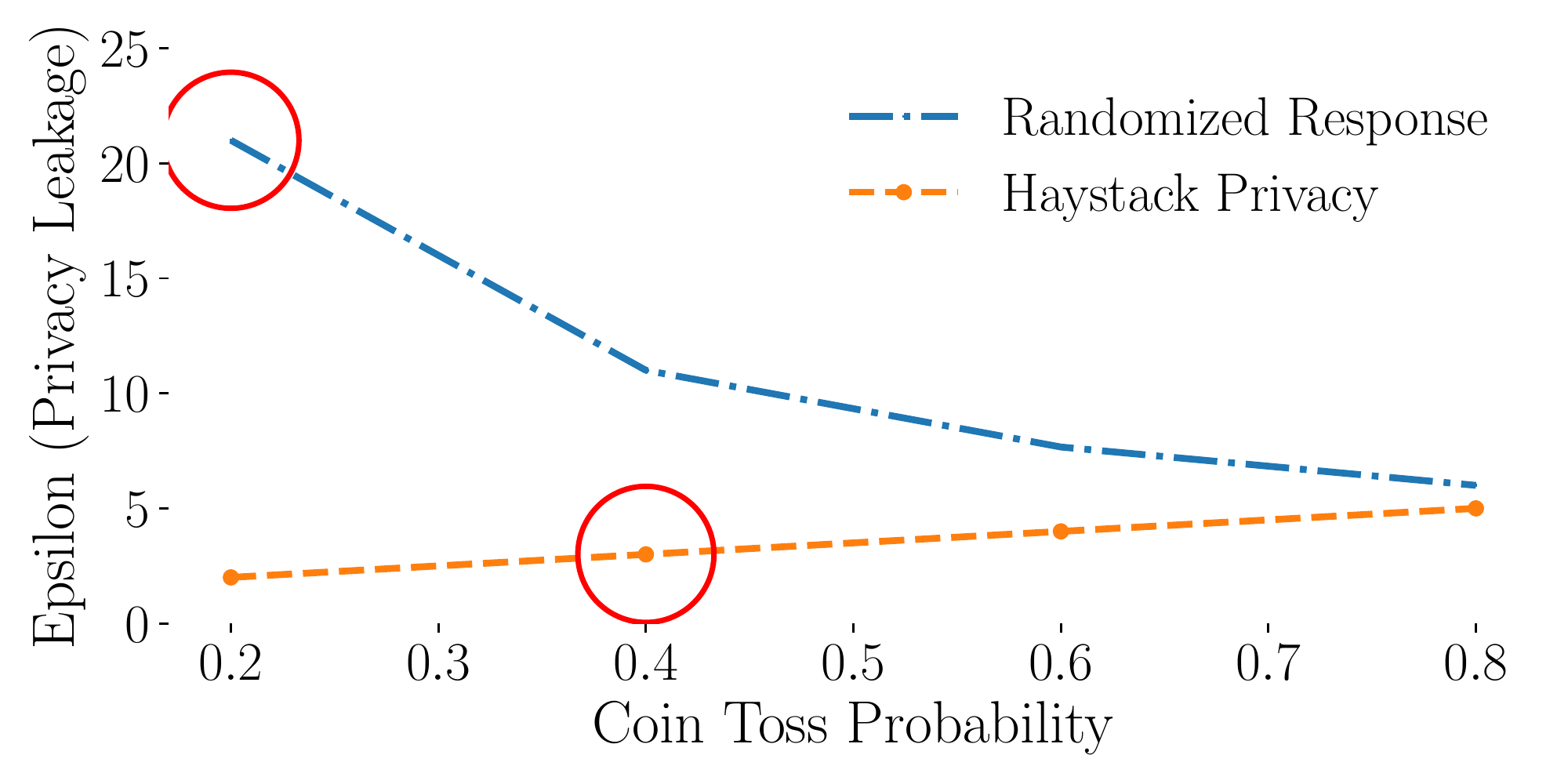}
    \caption{\textbf{(Privacy Leakage)} \titlename compared with Randomized Response privacy leakage as the coin toss probability increases. Higher epsilon means more information is leaked.}
    \label{fig:epsiloncomparison}
\end{figure}

\begin{figure}[!htb]
\centering
    \includegraphics[width=1\columnwidth]{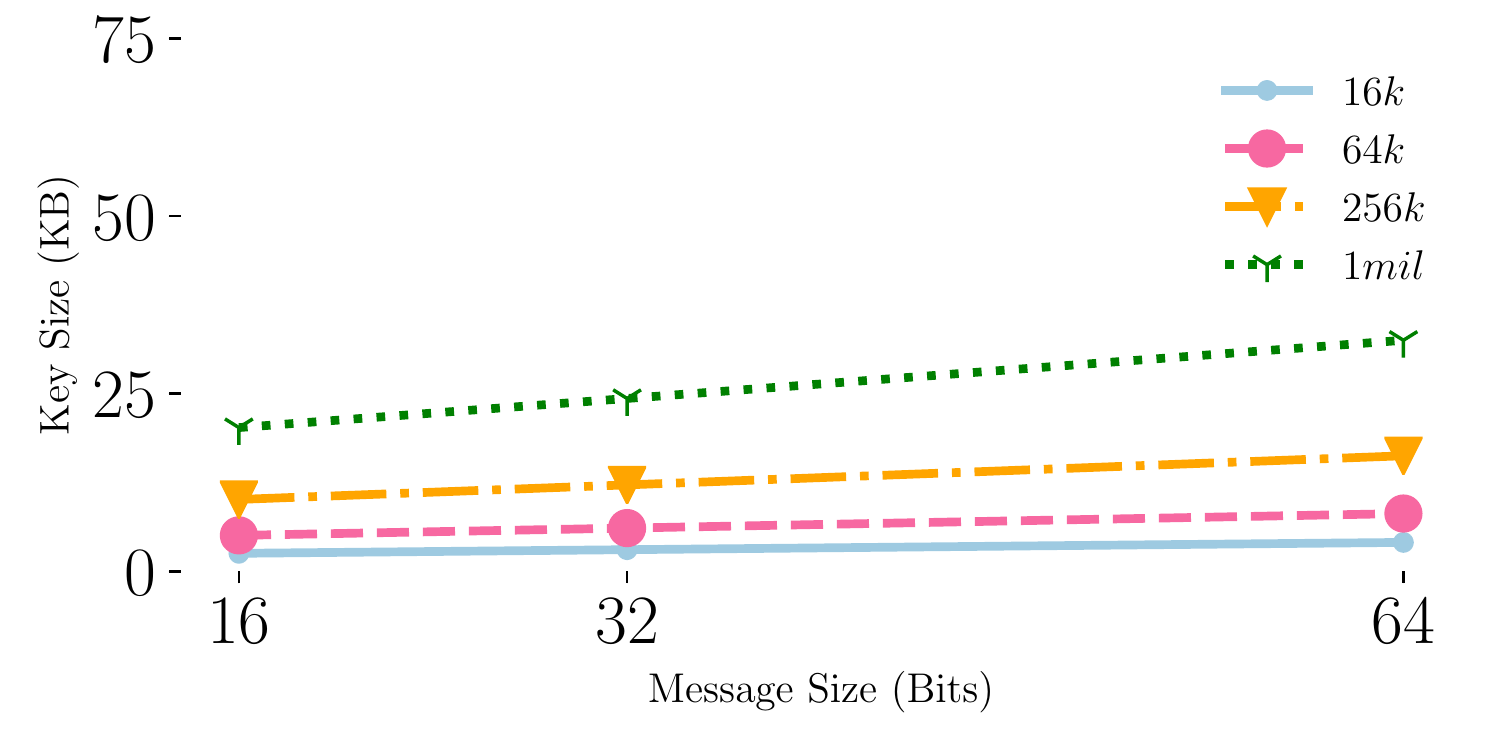}
    \caption{\textbf{FSS Keysize.} }
    \label{fig:keysize}    
\end{figure}

\begin{figure}[!htb]
\centering
    \includegraphics[width=1\columnwidth]{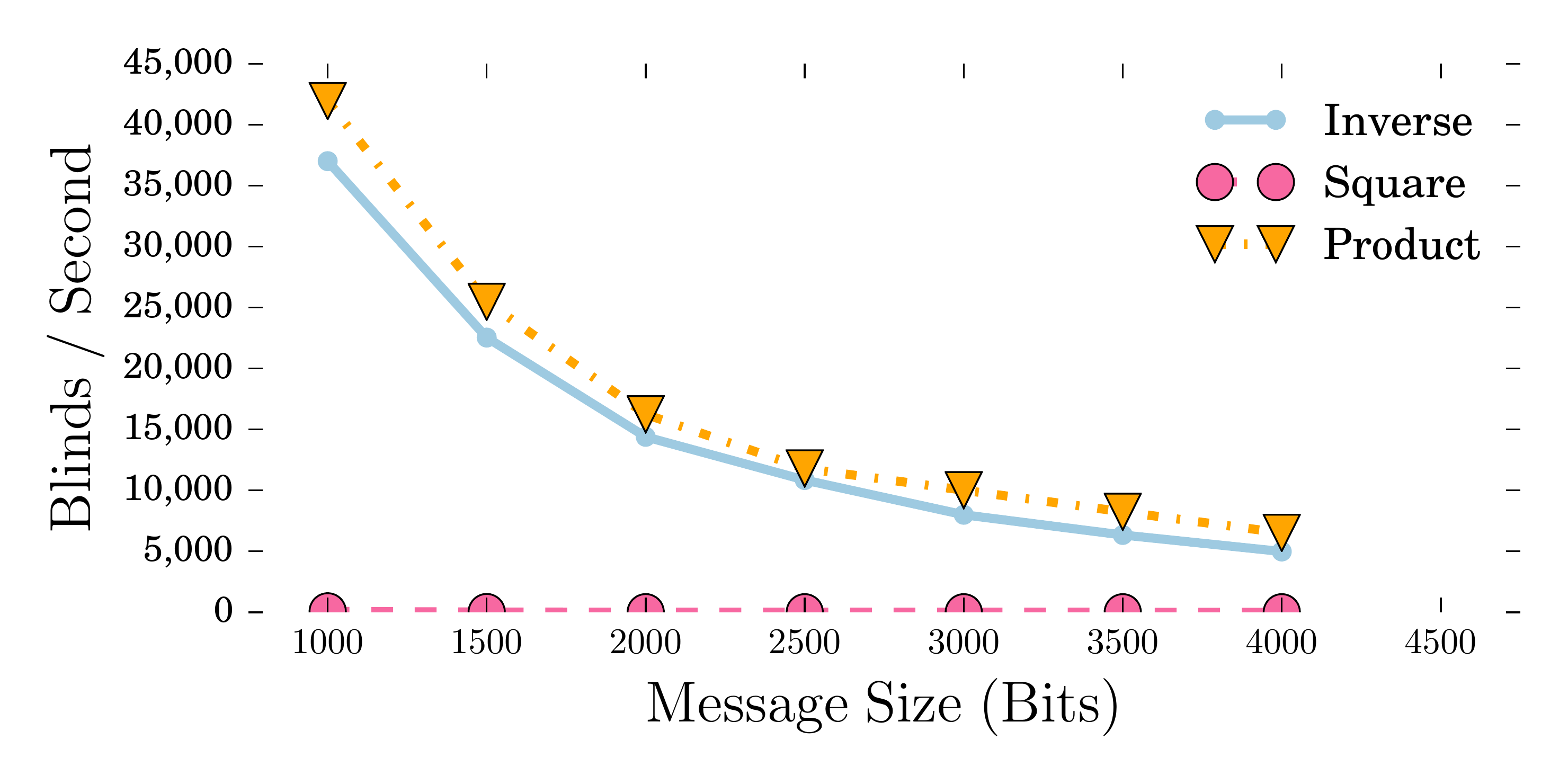}
    \caption{\textbf{MPC Verification Benchmark.} }
    \label{fig:verification-benchmark}
\end{figure}

\subsection{Privacy}
\label{sec:evaluation:privacy}

Figure~\ref{fig:epsiloncomparison} evaluates the privacy leakage comparing \titlename and the Randomized Response mechanism. \titlename uses the equation defined in  ~\ref{sec:dpprivacyguarantee} to measure the privacy leakage. The Randomized Response mechanism privacy leakage is defined in the Appendix~\ref{sec:randomizedresponseprivacy}.

The coin toss parameters used in Figure~\ref{fig:epsiloncomparison} has Randomized Response $flip1=0.8$ and $flip2=0.2$. \titlename has a sampling parameter of $0.45$. We could increase the value of the randomized response $flip2$ though the absolute error would grow even larger than show in Figure ~\ref{fig:pems}.

\subsection{Scalability}
\label{sec:evaluation:scalability}

~\\
\noindent \textbf{FSS Optimization} \label{sec:evaluation:fssoptimization} We evaluate our implementation of non-attributable writes on Amazon EC2 with c4.2xlarge instances to understand the impact of our optimization of the Function Secret Sharing primitive. 

We first perform a microbenchmark to evaluate the improvement of our optimization of evaluating shares to be performed by the aggregators. Figure~\ref{fig:fss-microbenchmark} shows our microbenchmark for the 3 party and 5 party case. Our microbenchmark shows several orders of magnitude improvement over of the default implementation.

We then evaluate the generation of shares. We evaluate the trade-off of key size by expanding a single seed  versus expanding multiple seeds with a smaller length. Figure~\ref{fig:genshares} shows the effect of the FSS optimization on the time to generate shares. We achieve close to half a reduction in share generation time.

We next evaluate our evaluation optimization on Amazon EC2. Figure~\ref{fig:fssscaling} shows the effect of applying the FSS optimization for the evaluation of the shares as described in Section~\ref{sec:fssoptimization}. We are able to achieve an order of magnitude improvement over the default implementation.

~\\
\noindent \textbf{Share Verification.} We now discuss the evaluation of our implementation of the FSS share verification~\cite{DBLP:conf/ccs/BoyleGI16}. The three algorithms for creating the blinding structure are ``square", ``product", and ``inverse" (see Appendix~\ref{sec:shareverificationexample} for more details).

Figure~\ref{fig:verification-benchmark} shows the scalability of the blinding operations. ``Product" is slightly faster than ``square" as ``product" must only do$(p-1)$ multiplications, while ``square" does $(p-1)$ exponent operations. ``Inverse" is the slowest as it performs $(p-1)$ multiplications and then a finite field inverse, where $p$ is the number of parties. The MPC verification performed by the aggregation servers is on the order of a couple hundred milliseconds and is extremely efficient.

 \section{Conclusion}

In this paper, we present the \titlename mechanism and demonstrate how to i) improve the privacy strength while preserving utility, ii) achieve scalable non-attributable writes, and iii) provide protection against pollution attacks whereby a single data owner may attempt to corrupt the entire database. To demonstrate its real-world applicability and practicality, the \titlename mechanism was implemented on Amazon's AWS cloud and shown to scalably achieve these properties. We believe this represents an important and timely advance towards open and shared Internet of Vehicles data.
 \section{Acknowledgements}

We thank the FSS Authors, Peter Reiher, Zengwen Yuan, and Torsten Braun for their helpful discussions.
 
\bibliographystyle{IEEEtranS}

\appendices

\section{Differential Privacy}
\label{sec:differentialprivacy}

Differential privacy has become the \emph{gold standard} privacy mechanism which ensures that the output of a sanitization mechanism does not violate the privacy of any individual inputs.  

\begin{definition}[\cite{DBLP:conf/icalp/Dwork06,DBLP:conf/tcc/DworkMNS06}]{($\epsilon$-Differential Privacy).}
\label{def:differentialprivacy}
A privacy mechanism $San()$ provides $\epsilon$-differential privacy if, for all datasets $D_1$ and $D_2$ differing on at most one record (i.e., the Hamming distance $H()$ is $H(D_1,D_2) \leq 1$), and for all outputs $O \subseteq Range(San())$:
\begin{equation}
\sup_{D_1,D_2}\frac{\Pr[San(D_1) \in O]}{\Pr[San(D_2) \in O]} \leq exp(\epsilon)
\label{eqn:dp}
\end{equation}
\end{definition}

That is, the probability that a privacy mechanism $San$ produces a given output is almost independent of the presence or absence of any individual record in the dataset.  The closer the distributions are (i.e., smaller $\epsilon$), the stronger the privacy guarantees become and vice versa. That is, a larger $\epsilon$ means that the two dataset distribution are far apart and leaks more information. A single record will induce distinguishable output fluctuations. We desire smaller $\epsilon$ values to induce $\epsilon$ \textit{indistinguishability}.

\subsection{Randomized Response Privacy Guarantee}
\label{sec:randomizedresponseprivacy}

\subsubsection{Privacy Guarantee of Randomized Response}
\label{sec:priv_g}
The randomized response mechanism achieves $\epsilon$-differential privacy, where:

\small
\begin{multline*}
\epsilon = \max \bigg({\ln \Big(\frac{\Pr[\textrm{Resp=`Yes' \textbar `Yes'}]} {\Pr[\textrm{Resp=`Yes' \textbar `No'}]} \Big), \ln \Big(\frac{\Pr[\textrm{Resp=`Yes' \textbar `No'}]} {\Pr[\textrm{Resp=`Yes' \textbar `Yes'}]}\Big)} \bigg)
\end{multline*}
\normalsize

More specifically, the randomized response mechanism~\cite{fox1986randomized} achieves $\epsilon$-differential privacy, where:
\begin{equation}
\label{eqn:e-forced}
\epsilon = \ln \Big( \frac{\pi_1 + (1 - \pi_1) \times \pi_2}{(1 - \pi_1) \times \pi_2} \Big)
\end{equation}

That is, if a data owner has the sensitive attribute $A$, then the randomized answer will be ``Yes'' with the probability of `$\pi_1 + (1 - \pi_1) \times \pi_2$'.  Else, if a data owner does not have the sensitive attribute, then the randomized answer will become ``Yes'' with the probability of `$(1 - \pi_1) \times \pi_2$'.
 \section{Share Verification}
\label{sec:shareverificationexample}

We now describe the MPC protocol~\cite{DBLP:conf/ccs/BoyleGI16} run amongst the aggregator parties to verify all data owner shares. The protocol does not violate data owner privacy and is extremely efficient as it does not utilize any publick-key primitives and relies solely on finite field operations.

We first describe the MPC protocol in detail and then provide an example.

\textbf{MPC Protocol}
Let $p$ represent the number of parties participating in the protocol.

Let $n$ represent the unit vector length (e.g., length of the bit string or number of database slots).

Let $m$ represent the number of bits of the message $M$. Let $M \in \mathbb{F}_Z$ where $Z$ is a relatively large prime number. \\

Given $\mathbb{F}_Z$ a finite field of characteristic $Z$ where $Z$ is a relatively large prime, let $\mathbf{R}$ be a blinding (randomization) matrix where the the values in the first row are chosen uniformly at random over ${0,...,Z-1}$.

This is a particular randomization matrix such that elements of each row is raised to the power of the first row, where the power is equivalent to the row number. There will be a total of $p$ rows, one for each party. That is, 

\begin{equation}
\mathbf{R}={\begin{bmatrix}
r_1 & r_2  & ...  & r_n \\ 
r_1^2 & r_2^2  & ...  & r_n^2 \\ 
... & ... & ... & ... \\
r_1^p & r_2^p  & ...  & r_n^p \\  	
\end{bmatrix}}
\end{equation}

We wish to secretly share a unit vector and verify that the shares correctly sum to the unit vector.

For example,

\begin{equation}
\hat{\textbf{u}}=\begin{bmatrix}
0\\ 
$M$\\ 
...\\
0
\end{bmatrix}
\end{equation}

The value can be $m$ bits taking on a value from the finite field of character of characteristic $p$ where $p$ is a relatively large prime.

To share  $\hat{\mathbf{u}}$, the user can randomly generate a total $p$ vectors {$\mathbf{V_i}$ 

\begin{equation}
\mathbf{V_i}=\begin{bmatrix}
v_{i,1}\\
v_{i,2}\\
...\\
v_{i,n}\\
\end{bmatrix}
\end{equation}

such that

\begin{equation}
\sum\limits_{i=1}^p \mathbf{V_i} = \mathbf{\hat{u}}
\end{equation}

We then blind these values such that

\begin{equation}
\sum\limits_{i=1}^p \mathbf{R} \cdot \mathbf{V_i} = \mathbf{R} \cdot \mathbf{\hat{u}}
\end{equation}

Let's describe an example where $n=2$ and $p=3$.

We know that sum of the vectors should equal the unit vector.
\begin{equation}
\begin{bmatrix}
v_{1,1}\\
v_{1,2}\\
\end{bmatrix}
+
\begin{bmatrix}
v_{2,1}\\
v_{2,2}\\
\end{bmatrix}
+
\begin{bmatrix}
v_{3,1}\\
v_{3,2}\\
\end{bmatrix}
=
\mathbf{\hat{u}}
\label{equation:unitvectorsum}
\end{equation}

We now apply the randomization (blinding) matrix.

\begin{align}
\begin{split}
\renewcommand\arraystretch{1.5}
\begin{bmatrix}
r_1 & r_2  \\ 
r_1^2 & r_2^2  \\ 
r_1^3 & r_2^3   \\
\end{bmatrix}
\cdot
\begin{bmatrix}
v_{1,1}\\
v_{1,2}\\
\end{bmatrix}
+
\begin{bmatrix}
r_1 & r_2  \\ 
r_1^2 & r_2^2  \\ 
r_1^3 & r_2^3   \\
\end{bmatrix}
\cdot
\begin{bmatrix}
v_{2,1}\\
v_{2,2}\\
\end{bmatrix}
+ \\
\begin{bmatrix}
r_1 & r_2  \\ 
r_1^2 & r_2^2  \\ 
r_1^3 & r_2^3   \\
\end{bmatrix}
\cdot
\begin{bmatrix}
v_{3,1}\\
v_{3,2}\\
\end{bmatrix}
=
\mathbf{R} \cdot \mathbf{\hat{u}}
\end{split}
\end{align}

\begin{align}
\begin{split}
\renewcommand\arraystretch{1.5}
\begin{bmatrix}
r_1 \cdot v_{1,1} + r_2 \cdot v_{1,2} \\
r^2_1 \cdot v_{1,1} + r^2_2 \cdot v_{1,2} \\
r^3_1 \cdot v_{1,1} + r^3_2 \cdot v_{1,2} \\
\end{bmatrix}
+
\begin{bmatrix}
r_1 \cdot v_{2,1} + r_2 \cdot v_{2,2} \\
r^2_1 \cdot v_{2,1} + r^2_2 \cdot v_{2,2} \\
r^3_1 \cdot v_{2,1} + r^3_2 \cdot v_{2,2} \\
\end{bmatrix}
+ \\
\begin{bmatrix}
r_1 \cdot v_{3,1} + r_2 \cdot v_{3,2} \\
r^2_1 \cdot v_{3,1} + r^2_2 \cdot v_{3,2} \\
r^3_1 \cdot v_{3,1} + r^3_2 \cdot v_{3,2} \\
\end{bmatrix}
=
\mathbf{R} \cdot \mathbf{\hat{u}}
\end{split}
\end{align}

\begin{equation}
\renewcommand\arraystretch{1.5}
\begin{bmatrix}
r_1~(v_{1,1}+v_{2,1}+v_{3,1}) + r_2~(v_{1,2}+v_{2,2}+v_{3,2}) \\ 
r^2_1~(v_{1,1}+v_{2,1}+v_{3,1}) + r^2_2~(v_{1,2}+v_{2,2}+v_{3,2}) \\ 
r^3_1~(v_{1,1}+v_{2,1}+v_{3,1}) + r^3_2~(v_{1,2}+v_{2,2}+v_{3,2}) \\ 
\end{bmatrix}\
=
\mathbf{R} \cdot \mathbf{\hat{u}}
\end{equation}

Since the summation of the elements of a unit vector should sum to zero, we can denote the value as follows

\begin{equation}
\begin{bmatrix}
a+b \\
a^2+b^2 \\
a^3+b^3 \\
\end{bmatrix}
=
\mathbf{R} \cdot \mathbf{\hat{u}}
\end{equation}

From Equation~\ref{equation:unitvectorsum} that the sum of the vectors is the unit vector. Thus, we then know that if the shares are properly formed that $a$ and $b$ should represent either all zeros or the blinded message. Thus, $(a+b)^2 - (a^2+b^2) =0$ and $(a+b)^3 - (a^3+b^3)=0$. 

If $a$ and $b$ are both zero then the terms fall out.

In the case of only $a$ or $b$ being the blinded message the terms fall out. 

If both $a$ and $b$ are non-zero then the difference will be a non-zero value. These shares are invalid and should be discarded.

\subsection{Alternate Algorithms}
There are two alternate algorithms for the ``square'' algorithm described above, which were also presented in~\cite{DBLP:conf/ccs/BoyleGI16}. The same process is used, but the structure of the blidning matrix is different, as well as the final check of $\mathbf{R} \cdot \mathbf{\hat{u}}$. The first algorithm is the ``product'' algorithm where 

\begin{equation}
\mathbf{R}={\begin{bmatrix}
r_{1,1} & r_{2,1}  & ...  & r_{n,1} \\ 
r_{1,2} & r_{2,2}  & ...  & r_{n,2} \\ 
... & ... & ... & ... \\
r_{1,p} & r_{2,p}  & ...  & r_{n,p} \\  	
\end{bmatrix}}
\end{equation}

such that

\begin{equation}
\forall i~\prod \limits_{j=1}^{p-1} r_{i,j} = r_{i,p}
\end{equation}

Then, we can apply the bliding matrix to our vectors  {$\mathbf{V_i}$, to achieve the final result:

\begin{equation}
\begin{bmatrix}
a_1+b_1 \\
a_2+b_2 \\
a_3+b_3 \\
\end{bmatrix}
=
\mathbf{R} \cdot \mathbf{\hat{u}}
\end{equation}

where

\begin{equation}
\prod \limits_{i=1}^{p-1} (a_i + b_i) = a_p + b_p
\end{equation}

An Alternative scheme is the ``inverse'' algorithm, which has a blinding matrix of 

\begin{equation}
\mathbf{R}={\begin{bmatrix}
r_{1,1} & r_{2,1}  & ...  & r_{n,1} \\ 
r_{1,2} & r_{2,2}  & ...  & r_{n,2} \\ 
... & ... & ... & ... \\
r_{1,p} & r_{2,p}  & ...  & r_{n,p} \\  	
\end{bmatrix}}
\end{equation}

such that

\begin{equation}
\forall i~\prod \limits_{j=1}^{p} r_{i,j} = 1
\end{equation}

Then,

\begin{equation}
\begin{bmatrix}
a_1+b_1 \\
a_2+b_2 \\
a_3+b_3 \\
\end{bmatrix}
=
\mathbf{R} \cdot \mathbf{\hat{u}}
\end{equation}

where

\begin{equation}
\prod \limits_{i=1}^{p} (a_i + b_i) = 1
\end{equation}

\subsection{Share Verification Analysis}
Here we analyze the protocol to ensure that data owners' responses are correctly formed unit vectors where all indexes are zero except for only one index.

\textbf{Correctness}  The protocol outputs whether the final answer is a unit vector (i.e., all the indexes are zero except for one location). If the vector is all zeroes then the sum will be zero. If the answer is a unit vector then the blinded message terms fall out leaving zero. If the vector is not a unit vector, the sum will be non-zero and we can discard this share.

\textbf{Privacy} All parties only view their own input and the final output. The blinding mechanism effectively masks the data owners true value.

\textbf{Fairness} All parties which participate will all view the same final answer as the shares sum to the same value.
 
\end{document}